\documentclass[%
 reprint,
superscriptaddress,
 amsmath,amssymb,
 aps,
]{revtex4-2}

\usepackage{graphicx}
\usepackage{dcolumn}
\usepackage{bm}
\usepackage{siunitx}
\DeclareSIUnit\angstrom{\protect \text {Å}}
\usepackage{svg}
\usepackage{xr}
\usepackage{braket}
\usepackage{color, colortbl}
\usepackage{comment}

\newcommand{\blue}[1]{\textcolor{blue}{#1}} 

\begin{document}

\preprint{APS/123-QED}

\title{Combining experiments on luminescent centres in hexagonal boron nitride with the polaron model and \emph{ab initio} methods towards the identification of their microscopic origin}

\author{Moritz Fischer}
 \affiliation{Department of Electrical and Photonics Engineering, Technical University of Denmark, 2800 Kgs. Lyngby, Denmark}
 \affiliation{Centre for Nanostructured Graphene, Technical University of Denmark, 2800 Kgs. Lyngby, Denmark}
 \affiliation{NanoPhoton - Center for Nanophotonics, Technical University of Denmark, 2800 Kgs. Lyngby, Denmark}
\author{Ali Sajid}
 \affiliation{Centre for Nanostructured Graphene, Technical University of Denmark, 2800 Kgs. Lyngby, Denmark}
 \affiliation{Department of Physics, Technical University of Denmark, 2800 Kgs. Lynby, Denmark}
\author{Jake Iles-Smith}
 \affiliation{Department of Electrical and Electronic Engineering, The University of Manchester, Sackville Street Building, Manchester M1 3BB, United Kingdom}
\author{Alexander H\"otger}
 \affiliation{Walter Schottky Institute and Physics Department, Technical University of Munich, 85748 Garching, Germany}
\author{Denys I. Miakota} 
 \affiliation{Department of Electrical and Photonics Engineering, Technical University of Denmark, 2800 Kgs. Lyngby, Denmark}
\author{Mark K. Svendsen}
 \affiliation{Department of Physics, Technical University of Denmark, 2800 Kgs. Lynby, Denmark}
\author{Christoph Kastl}
 \affiliation{Walter Schottky Institute and Physics Department, Technical University of Munich, 85748 Garching, Germany}
\author{Stela Canulescu}
\affiliation{Department of Electrical and Photonics Engineering, Technical University of Denmark, 2800 Kgs. Lyngby, Denmark}
\author{Sanshui Xiao}
 \affiliation{Department of Electrical and Photonics Engineering, Technical University of Denmark, 2800 Kgs. Lyngby, Denmark}
 \affiliation{Centre for Nanostructured Graphene, Technical University of Denmark, 2800 Kgs. Lyngby, Denmark}
 \affiliation{NanoPhoton - Center for Nanophotonics, Technical University of Denmark, 2800 Kgs. Lyngby, Denmark}
\author{Martijn Wubs}
 \affiliation{Department of Electrical and Photonics Engineering, Technical University of Denmark, 2800 Kgs. Lyngby, Denmark}
 \affiliation{Centre for Nanostructured Graphene, Technical University of Denmark, 2800 Kgs. Lyngby, Denmark}
 \affiliation{NanoPhoton - Center for Nanophotonics, Technical University of Denmark, 2800 Kgs. Lyngby, Denmark}
\author{Kristian S. Thygesen}
 \affiliation{Centre for Nanostructured Graphene, Technical University of Denmark, 2800 Kgs. Lyngby, Denmark}
 \affiliation{Department of Physics, Technical University of Denmark, 2800 Kgs. Lynby, Denmark}
\author{Alexander W. Holleitner}
 \affiliation{Walter Schottky Institute and Physics Department, Technical University of Munich, 85748 Garching, Germany}
\author{Nicolas Stenger}
 \email[Corresponding author. Email: ]{niste@dtu.dk}
 \affiliation{Department of Electrical and Photonics Engineering, Technical University of Denmark, 2800 Kgs. Lyngby, Denmark}
 \affiliation{Centre for Nanostructured Graphene, Technical University of Denmark, 2800 Kgs. Lyngby, Denmark}
 \affiliation{NanoPhoton - Center for Nanophotonics, Technical University of Denmark, 2800 Kgs. Lyngby, Denmark}

\newpage
\begin{abstract}

The two-dimensional material hexagonal boron nitride (hBN) hosts luminescent centres with emission energies of~$\sim\SI{2}{eV}$ which exhibit pronounced phonon sidebands.
We investigate the microscopic origin of these luminescent centres by combining \emph{ab initio} calculations with non-perturbative open quantum system theory to study the emission and absorption properties of 26 defect transitions. 
Comparing the calculated line shapes with experiments 
we narrow down the microscopic origin to three carbon-based defects: $\mathrm{C_2C_B}$, $\mathrm{C_2C_N}$, and $\mathrm{V_NC_B}$.
The theoretical method developed enables us to calculate so-called photoluminescence excitation (PLE) maps, which show excellent agreement with our experiments.
The latter resolves higher-order phonon transitions, thereby confirming both the vibronic structure of the optical transition and the phonon-assisted excitation mechanism with a phonon energy~$\sim\SI{170}{meV}$. 
We believe that the presented experiments and polaron-based method accurately describe luminescent centres in hBN and will help to identify their microscopic origin.

\end{abstract}

\maketitle

\section*{Introduction}
Luminescent centres in hexagonal boron nitride (hBN) have gained an increased scientific interest due to the demonstration of single photon emission with a brightness comparable to semiconductor quantum dots~\cite{SolidStateSPEsReview2016}. These luminescent centres emit at photon energies around $\SI{2}{eV}$ and persist even at room temperature~\cite{TranNanoTech2016}, making hBN a promising material to realise future optoelectronic technologies such as quantum telecommunication~\cite{PhotonicQuantumTechnology2009,QKDAharonovich2023} and quantum sensing~\cite{QuantumSensing2017}.\par

The microscopic origin of $\SI{2}{eV}$ luminescent centres in hBN has been experimentally narrowed down to carbon-based defects by bottom-up and post-growth techniques~\cite{MendelsonCarbon2021}, however the atomic structure of the underlying defect remains elusive.
One subset of $\SI{2}{eV}$ luminescent centres are group~I centres~\cite{FischerSciAdv2021} with pronounced phonon side bands (PSB). 
The chemical composition of the defect will naturally alter the mechanical vibrations of the crystal which will in-turn modify the structure of the PSB observed in photoluminescence. This motivates the comparison of experimental lineshapes with those obtained via 
\emph{ab initio} methods~\cite{JaraC2CN2021,LiSmartPing2022,SajidVNCB2020}.

A typical approach to calculating the photoluminescence of defect transitions is to first calculate the electronic structure and phonon modes using \emph{ab initio} methods, before using this information to calculate the photoluminescence spectrum with the generating function approach~\cite{TheoryTawfik2017,SajidVNCB2020,JaraC2CN2021,LiSmartPing2022,TheoryLinderalv2021,Auburger2021,Skoff2023}. 
Whilst this method can calculate the linear absorption and emission spectra with atomistic precision, it cannot resolve coherent dynamics between electronic states of the defect transition, and is limited to unstructured electromagnetic environments.
As the field strives towards coherent control of hBN luminescent centres~\cite{WiggerCoherent2022}, and interfacing emitters with plasmonic~\cite{NguyenGold2018} and photonic structures~\cite{VoglCavity2019}, new theoretical methods are required to describe the behaviour of defect complexes.

In this joint experiment-theory investigation, we study the photoluminescence and photoluminescence emission (PLE) of $\SI{2}{eV}$ luminescent centres in hBN.
We combine a non-perturbative master equation treatment of electron-phonon interaction with \emph{ab initio} methods to calculate the photoluminescence line shapes of 26 candidate defect transitions. 
This methodology, based on the polaron formalism~\cite{Nazir_PolaronMethodReview2016}, maintains the atomistic accuracy of the generating function approach, 
while allowing to simulate PLE maps of the studied defect transition by taking into account external driving fields.
By comparing the calculated photolumiscence against measurements for twelve group~I centres we exclude all but three candidate defects: the neutral substitutional carbon trimers $\mathrm{C_2C_B}$ and $\mathrm{C_2C_N}$ as well as $\mathrm{V_NC_B}$.
Here, $\mathrm{C_2C_B}$ and $\mathrm{C_2C_N}$ are shorthand notations of $\mathrm{C_BC_NC_B}$ and $\mathrm{C_NC_BC_N}$, respectively.
By comparing measured and theoretical PLE, which may be calculated directly using the polaron formalism, we are able to resolve higher-order phonon transitions.
By studying the emission of zero-phonon line (ZPL) and PSB simultaneously, we confirm that the excitation mechanism in group~I emitters is phonon-assisted with a phonon energy of $\sim\SI{170}{meV}$.
The combination of theoretical and experimental methods presented here will help to identify the microscopic origin of luminescent centres in hBN.\par



\section*{Results and discussion}
To generate luminescent centres in hBN we use a process that was developed recently in our group~\cite{FischerSciAdv2021}. Briefly, we irradiate high-quality multilayer hBN with oxygen atoms and use subsequent annealing to achieve high densities of luminescent centres emitting at photon energies around $\SI{2}{eV}$.
In our previous work, we carried out room-temperature characterisation of group~I and group~II centres, as defined in Ref.~\cite{FischerSciAdv2021}. In the work presented here, we focus our attention on low-temperature characterisation of group~I centres which show a specific line shape with pronounced PSB at ZPL detunings around $\SI{170}{meV}$, as observed also in other works~\cite{FeldmanCrosscorrelation2019,BollPhotophysics2020,Martinez_Bleaching2016}. Furthermore, these group~I centres emit at photon energies around $\SI{2}{eV}$ just like single-photon emitters in hBN~\cite{TranNanoTech2016}.
In this work, we define luminescent centres as experimental line shapes while theoretical line shapes are called defect transitions. Single-photon emitters are luminescent centres with an auto-correlation function fulfilling the criterion $g^{(2)}(0)<\SI{0.5}{}$. 
In the following we will study luminescent centres with a focus on group~I line shapes.\par

\begin{figure*}
      \includegraphics[width=\textwidth]{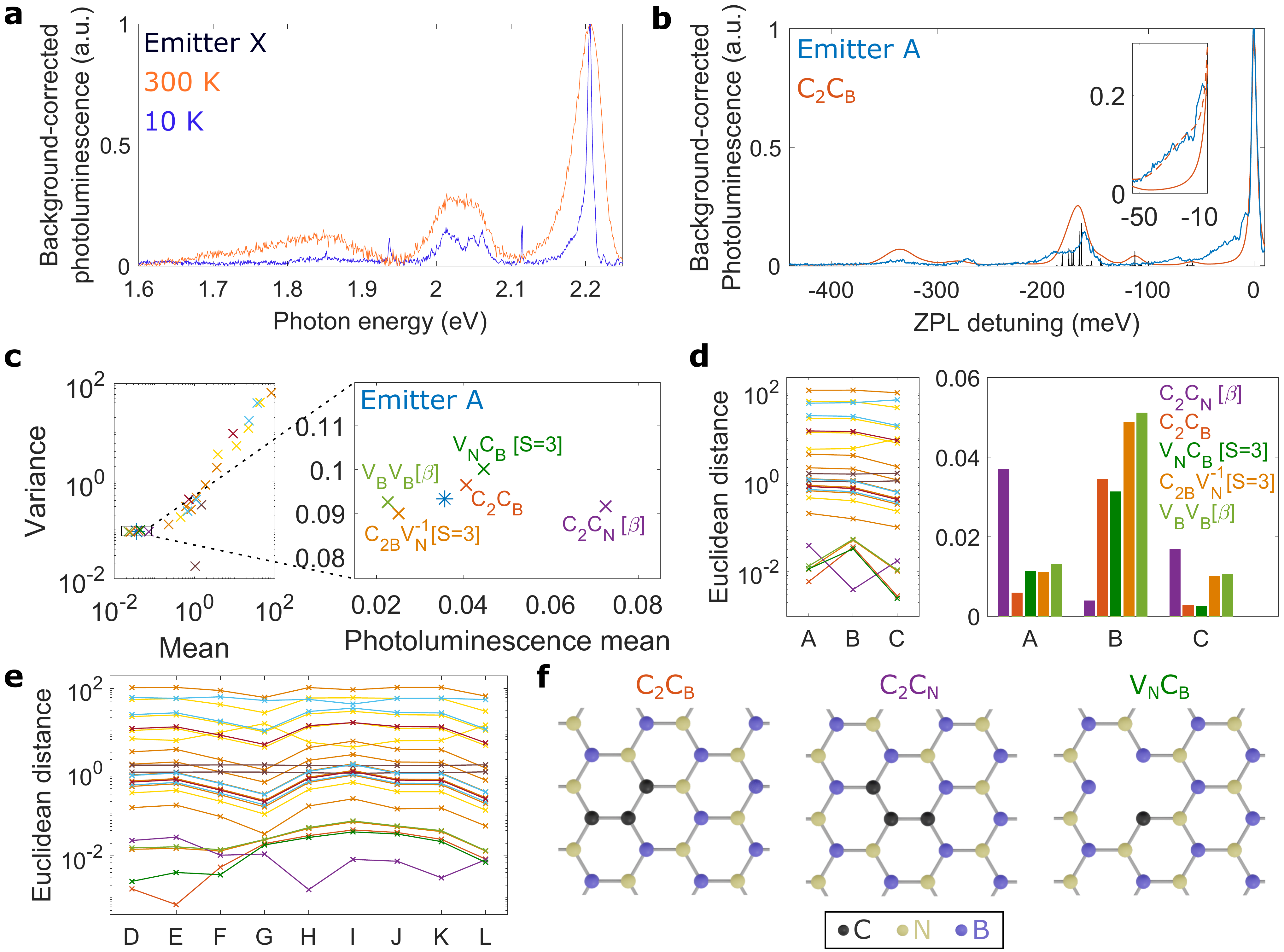}
      \caption{\label{fig1:Scatter}\textbf{Screening of defect transitions.}
      (a)~Room- and low-temperature spectra of Emitter~X under $\SI{2.37}{eV}$ excitation. a.u., arbitrary units.
      (b)~Comparison of low-temperature photoluminescence of Emitter~A (under $\SI{2.321}{eV}$ excitation) with the theoretical emission line shape of $\mathrm{C_2C_B}$ using the PBE functional (without acoustic phonons). The black bars show the partial Huang-Rhys factors (a.u.) obtained with the PBE functional.
      The inset shows the theoretical line shape of $\mathrm{C_2C_B}$ using the HSE06 functional with and without acoustic phonons in dashed and solid lines, respectively.
      (c)~Scatter plot comparing the photoluminescence mean and variance of theoretically calculated defect line shapes with the experimental data of Emitter~A. The most likely defect transitions are shown in the right panel which is the region close to the experimental data, marked by the square in the left panel.
      (d)~Euclidean distances of all defect transitions in corresponding colors to~(c) for Emitter~A,~B and~C. The histogram on the right shows the distances for the five most likely defect candidates.
      (e)~Euclidean distances for Emitter~D to~L in corresponding colors to~(c). All emitters show small distances for the same five defect transitions.
      (f)~Schematics of $\mathrm{C_2C_B}$, $\mathrm{C_2C_N}$, and $\mathrm{V_NC_B}$ where the latter is taken from Ref.~\cite{Sajid_PRB_VNCB2018}.
      All experimental photoluminescence line shapes are background-corrected as outlined in Supplementary Information~\ref{subsec:SI_BackgroundAndPSB}.
      We note that the PBE functional is used for all defect transitions (including $\mathrm{C_2C_B}$, $\mathrm{C_2C_N}[\beta]$, and $\mathrm{V_NC_B}[S=3]$), except for the inset in~(b) where the HSE06 functional is used.}
\end{figure*}

Fig.~\ref{fig1:Scatter}a shows the photoluminescence of a group~I centre at room and low temperature. Decreasing the temperature results in a narrower ZPL and reveals a detailed structure of the PSB. When comparing photoluminescence line shapes for different group~I centres at low temperature, 
we find variations among the PSB line shapes as shown in Supplementary Information~\ref{subsec:SI_FurtherGroupISpectra}.
These differences correspond to variations in the electron-phonon coupling of the defect transition among the luminescent centres, and may be due to different underlying defects. 
It is therefore necessary to compare several different defect transitions to \emph{individual} luminescent centres, as outlined below. 
This is in contrast to comparing total Huang-Rhys factors of several defect transitions with experiments, which neglects the detailed information about the spectral structure of the PSB.\par

\textbf{Screening of defect transitions}\\
To get insight into the microscopic origin of $\SI{2}{eV}$ luminescent centres in hBN, we study 26 different defect transitions. 
We focus on carbon-based defects as these have been experimentally demonstrated to be responsible for luminescent centers emitting around~$\SI{2}{eV}$~\cite{MendelsonCarbon2021}. Furthermore, our generation process can easily generate carbon-based defects by incorporating ubiquitous hydrocarbons from the annealing environment as well as by carbon impurities intrinsically present in the hBN~\cite{FischerSciAdv2021}. 
We note that we use the PBE functional for this screening of defect transitions while the HSE06 functional is used for a more accurate study of a selected subset of defect transitions.\par

To obtain the theoretical photoluminescence line shapes, we use \emph{ab initio} methods to calculate the ZPL energies and partial Huang-Rhys factors of all studied defect transitions (see Supplementary Information~\ref{subsec:SI_Fig3TheoryDetails}). With these Huang-Rhys factors at hand, we can construct the spectral density~\cite{kamper2022signatures}
\begin{align}
 J_\mathrm{Sim}(\omega) = \sum_k s_k\delta(\omega_k - \omega)
 \label{eq:SpectralDensityAbInitio}
\end{align}
Here, $\omega$ is the angular frequency and $s_k$ the partial Huang-Rhys factor associated to a phonon mode of angular frequency $\omega_k$. To account for the natural lifetime of phonons, we approximate the $\delta$-functions with Gaussian functions as outlined in Supplementary Information~\ref{subsec:JakeFitExperiments}.
For details on the \emph{ab initio} simulations, we refer the reader to the Methods.\par

With the spectral density at hand, we can calculate the dynamics and optical properties of the system using the polaron method. 
Here, the electronic states of the defect transition are dressed by vibrational modes of the phonon environment~\cite{Nazir_PolaronMethodReview2016,iles2017phonon}.
This enables one to derive a quantum master equation that is non-perturbative in the electron-phonon coupling strength, and thus captures phonon sideband processes in the photoluminescence~\cite{iles2017limits,iles2017phonon}. 
In Fig.~\ref{fig1:Scatter}b we compare the photoluminescence of Emitter~A against the line shape for the carbon trimer $\mathrm{C_2C_B}$, calculated using the polaron theory.  
The black bars show the partial Huang-Rhys factors ($s_k$) illustrating the origin of the phonon side bands from \emph{ab initio} methods. 
We highlight that the line shape of $\mathrm{C_2C_B}$ is unique for Emitter~A since we fit each theoretical line shape to \emph{individual} experimental line shapes, with the ZPL linewidth as the only free fitting parameter (see Supplementary Information~\ref{subsec:JakeFitExperiments}).\par

In order to screen the 26 different defect transitions calculated with \emph{ab initio} methods, 
we calculate the photoluminescence mean $\left\langle S \right\rangle$ and variance $\sigma$, defined as,
\begin{align*}
    \left\langle S \right\rangle = \frac{1}{N} \sum_{i=1}^N S(\Delta_i) ,~~ 
    \sigma^2 = {\frac{1}{N} \sum_{i=1}^N \left( S(\Delta_i) - \left\langle S \right\rangle \right)^2},
\end{align*}
where $S(\Delta_i)$ is the intensity of the spectrum at the ZPL detuning $\Delta_i$, and $N$ is the number of data points 
(see Methods). 
If the photoluminescence mean and variance of a defect transition are close to the experimental values, the theoretical photoluminescence line shape is similar to the experimental one, thus providing a coarse approach to screening relevant defect transitions.\par 

As an example, we apply this method to Emitter~A and plot the photoluminescence mean and variance for all calculated defect transitions along with the experimental values (Fig.~\ref{fig1:Scatter}c). 
Here, different colors correspond to different defects, some of which have several optically active transitions (details in Supplementary Information~\ref{subsec:DetailedScatterPlots}). The scatter plot (in double logarithmic scale) shows that the photoluminescence mean and variance for most defect transitions differ significantly from the experimental data and thus are unlikely candidates,
while the most likely defect transitions are clustered around the experimental data of Emitter~A, as shown in the right panel of Fig.~\ref{fig1:Scatter}c.\par

Inspired by work on Raman spectroscopy~\cite{EuclideanDistanceThygesen2020}
, we define an Euclidean distance $d$ to the experiment
\begin{align*}
    d = \sqrt{ (\left\langle S \right\rangle - \left\langle E \right\rangle)^2 + (\sigma - s_E )^2},
\end{align*}
where, $\left\langle E \right\rangle$ and $s_E$ are the experimental mean and variance while $\left\langle S \right\rangle$ and $\sigma$ are the mean and variance for each defect transition. 
For Emitter~A, the line shape of $\mathrm{C_2C_B}$ is very similar to the experiments, thus $\mathrm{C_2C_B}$ is located closely to the experiment in the scatter plot (Fig.~\ref{fig1:Scatter}c) and shows a small Euclidean distance (Fig.~\ref{fig1:Scatter}d in logarithmic scale).\par

The scatter plot shown in Fig.~\ref{fig1:Scatter}c depends on the luminescent centre studied, since the ZPL line width differs among luminescent centres. 
Therefore, we calculate an \emph{individual} scatter plot for each luminescent centre (for Emitter~B and~C these are given in Supplementary Information~\ref{subsec:DetailedScatterPlots}) with the resulting Euclidean distances shown along with Emitter~A in Fig.~\ref{fig1:Scatter}d.
We find that the same five defect transitions show similar photoluminescence line shapes for Emitter~A,~B and~C. 
Expanding our analysis to twelve group~I centres (line shapes shown in Supplementary Information~\ref{subsec:SI_FurtherGroupISpectra}), we find that for all experiments the five closest defect transitions are identical to the ones for Emitter~A (Fig.~\ref{fig1:Scatter}e). Therefore, we focus our attention on these five defect transitions while discarding the other 21 transitions.\par

\textbf{Second screening step}\\
The statistical analysis described above is independent on the theoretical ZPL energy because we match the theoretical line shapes with the experimental ZPL energies. Thus, comparing theoretical with experimental ZPL energies can help to narrow down the microscopic origin even further.
For twelve group~I centres the experimental ZPL energies are between $\SI{2.0}{}$ and $\SI{2.3}{eV}$ (see Tab.~\ref{tab:ZPLEnergiesSelectedDefects}) and thus theoretical ZPL energies much different to this range reveal less likely candidates.
Among the five defect transitions showing small Euclidean distances to these twelve centres (Fig.~\ref{fig1:Scatter}), we neglect $\mathrm{V_BV_B}[\beta]$ and $\mathrm{C_{2B}V_N^{-1}[S=3]}$ since their predicted ZPL energies are too high  and too low, respectively~(see Tab.~\ref{tab:ZPLEnergiesSelectedDefects}).
On the contrary, the ZPL energies of $\mathrm{C_2C_B}$, $\mathrm{C_2C_N}[\beta]$, and $\mathrm{V_NC_B}[S=3]$ are similar to the experiments as shown in Tab.~\ref{tab:ZPLEnergiesSelectedDefects}. In the following, we call these three defects $\mathrm{C_2C_B}$, $\mathrm{C_2C_N}$, and $\mathrm{V_NC_B}$. 
The ZPL energies of these three defects show good agreement with experiments, while even the relatively more accurate HSE06 functional shows slightly smaller energies compared to the experimental values.\par 

All in all, we can exclude 23 defect transitions for twelve group~I centres and identified three most likely candidates: $\mathrm{C_2C_B}$, $\mathrm{C_2C_N}$, and $\mathrm{V_NC_B}$. Fig.~\ref{fig1:Scatter}f shows the schematics of these three defect transitions. We note that $\mathrm{V_NC_B}$ needs a nearby defect to populate the ground state, because the intersystem crossing from the singlet state can be neglected at our experimental detunings (see Supplementary Information~\ref{subsec:SI_Fig3_VNCBLevelScheme}).\par

\setlength{\tabcolsep}{0pt}
\definecolor{tableColor}{rgb}{0.07, 0.62, 1}

\begin{table}
    \centering
    \begin{tabular}{lccc}
        \rowcolor{tableColor!70}
        ~Defect transition~              &~~PBE~~                &~~HSE06~~     & Experiments \\ \hline
        \rowcolor{tableColor!30}
        $\mathrm{C_2C_B}$               &~~$\SI{1.16}{eV}$~~&~~$\SI{1.36}{eV}$~~&$\SI{2.0}{}$~-~$\SI{2.3}{eV}$ \\
        \rowcolor{tableColor!40}
        $\mathrm{C_2C_N}[\beta]$        &~~$\SI{1.51}{eV}$~~&~~$\SI{1.67}{eV}$~~&\\
        \rowcolor{tableColor!30}
        $\mathrm{V_NC_B}[S=3]$          &~~$\SI{1.39}{eV}$~~&~~$\SI{1.75}{eV}$~~&\\
        \rowcolor{tableColor!40}
        $\mathrm{V_BV_B}[\beta]$        &~~$\SI{2.70}{eV}$~~&--&\\
        \rowcolor{tableColor!30}
        $\mathrm{C_{2B}V_N^{-1}[S=3]}$  &~~$\SI{0.90}{eV}$~~&--&\\
    \end{tabular}
    \caption{\label{tab:ZPLEnergiesSelectedDefects}\textbf{Theoretical and experimental ZPL energies.}
    The applied functional (PBE or HSE06) for the calculations of the defect transitions is given by the column name while the range of ZPL energies of Emitter~A to~L are shown in the rightmost column (details in Supplementary Information~\ref{subsec:SI_FurtherGroupISpectra}).
    We did not use the HSE06 functional for $\mathrm{V_BV_B}[\beta]$ and $\mathrm{C_{2B}V_N^{-1}[S=3]}$ since their PBE results are very different to the experimental ZPL energies.
    The spin minority channel is labelled by $\beta$ and $S=3$ denotes the triplet state while we suppress the notations for spin majority channel and other spin states (details in Supplementary Information~\ref{subsec:SI_Fig3TheoryDetails}).}    
\end{table}

\textbf{PLE spectroscopy}\\
In contrast to the generating function approach, our polaron formalism allows to introduce an external driving field in the system Hamiltonian, thereby directly simulating PLE experiments.
For the latter, the photoluminescence is measured as a function of laser detuning which is defined as the laser energy minus the ZPL energy of the studied luminescent centre.\par

\begin{figure}
  \includegraphics[width=.85\columnwidth]{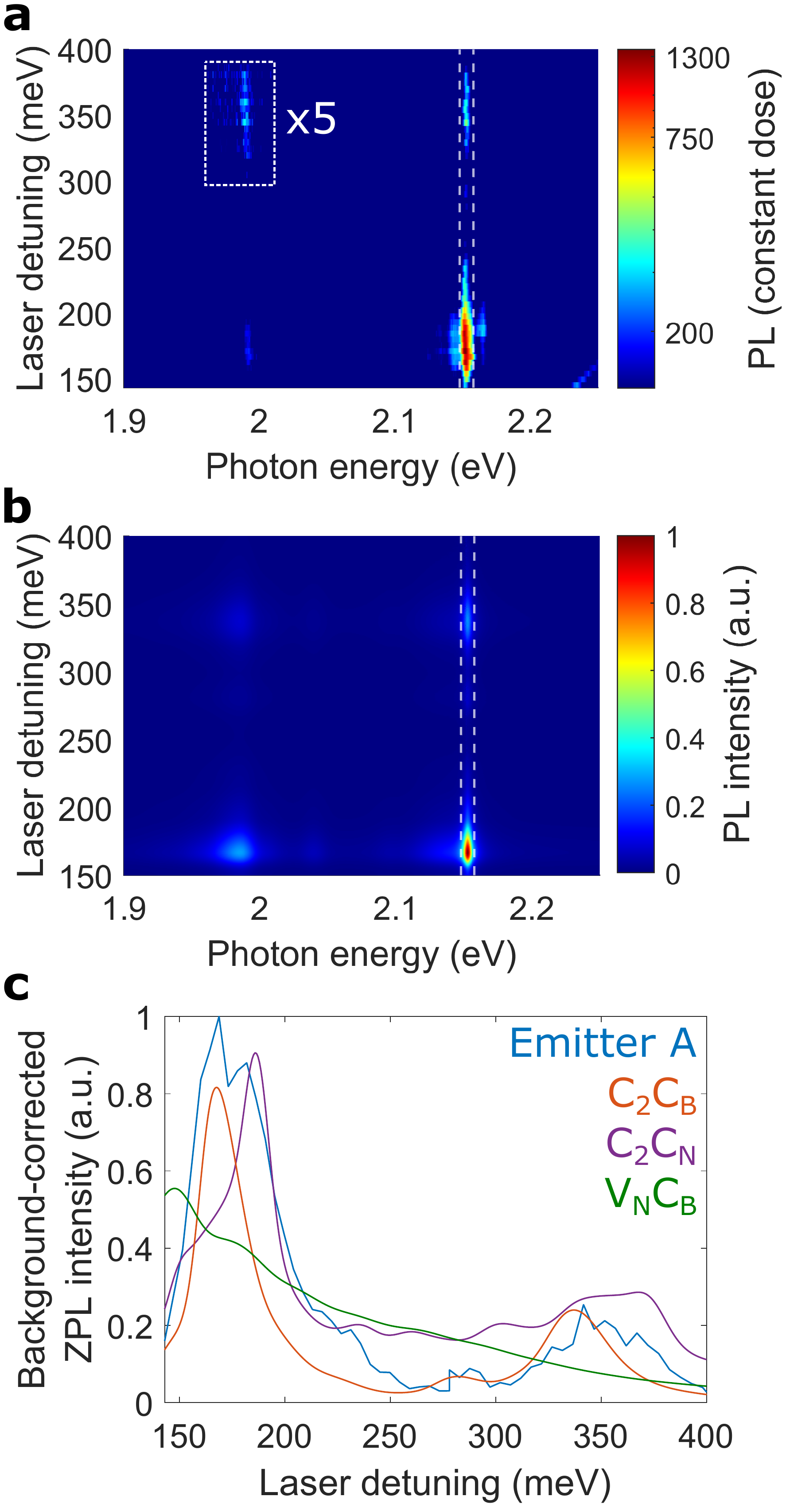}
  \caption{\label{fig2:PLE}\textbf{PLE spectroscopy.} (a)~Experimental PLE map of Emitter~A at $\SI{10}{K}$, background corrected as outlined in Supplementary Information~\ref{subsec:SI_Fig4ExtendedDetunings}. The photoluminescence (PL) in the highlighted box is multiplied by 5. The dim upward line in the bottom right corner corresponds to the silicon Raman~$\sim\SI{520}{cm^{-1}}$.
  (b)~Theoretical PLE map of $\mathrm{C_2C_B}$, shifted to the experimental ZPL energy of $\SI{2.153}{eV}$. Here, the HSE06 functional is used and acoustic phonons are included in the spectral density. 
  (c)~Comparison of the experimental ZPL intensity, indicated by dashed lines in~(a) and~(b), with $\mathrm{C_2C_B}$, $\mathrm{C_2C_N}$, and $\mathrm{V_NC_B}$. We note that the experimental ZPL intensity is rescaled to $\SI{170}{meV}$ and background-corrected as outlined in Supplementary Information~\ref{subsec:SI_Fig4ExtendedDetunings}. Furthermore, all theoretical curves include acoustic phonons as described in the main text.}
\end{figure}

Fig.~\ref{fig2:PLE}a shows the experimental PLE data (in logarithmic scale) of Emitter~A, i.e. the photoluminescence as a function of laser detuning (details and further group~I centres in Supplementary Information~\ref{subsec:SI_PLEDetails} and~\ref{subsec:SI_RescalingPLE}).
The ZPL emission at photon energies around $\SI{2.15}{eV}$ is strong at a detuning of $\SI{168}{meV}$ and shows intermediate strength at $\SI{341}{meV}$ detuning. Moreover, the PSB emission around $\SI{1.99}{eV}$ is enhanced at the same laser detunings. In particular, the enhanced PSB emission at $\SI{341}{meV}$ detuning (highlighted by the white box) shows strong evidence of a phonon-assisted mechanism, as outlined below.\par 

\begin{figure*}
  \includegraphics[width=0.85\textwidth]{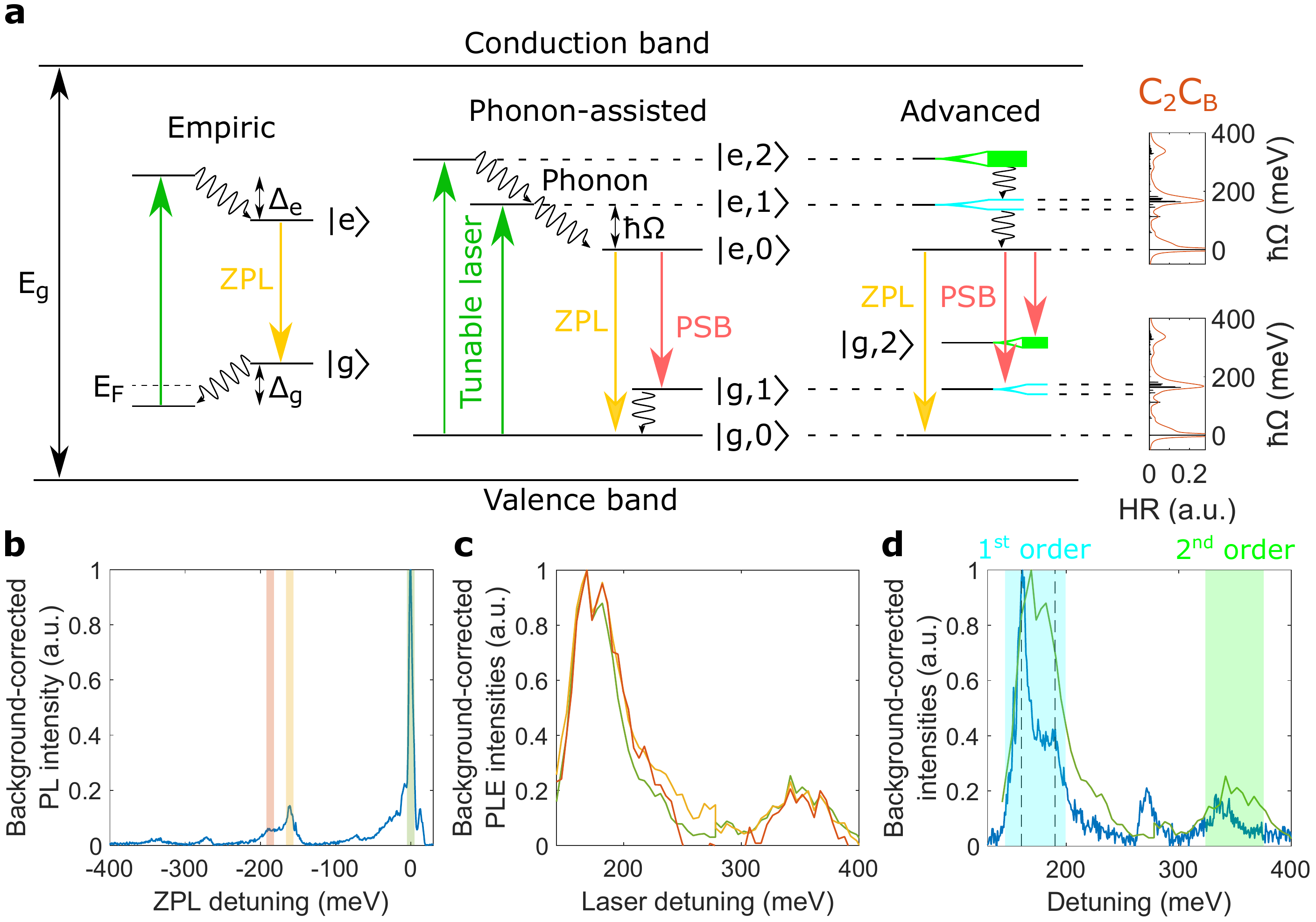}
  \caption{\label{fig3:ExcitationMechanism}\textbf{Excitation mechanism of Emitter~A.}
  (a)~Empiric, phonon-assisted, and advanced excitation mechanisms. The zero-phonon line (ZPL) and the phonon sidebands (PSB) are shown in yellow and red, respectively. The empiric mechanism has relaxation processes with energies $\Delta_e$ and $\Delta_g$ while the phonon-assisted mechanism shows relaxation via phonons with energy~$\hbar \Omega$. Here, the excited states are labelled with $\ket{e,n}$ corresponding to the electron in the excited state and its environment occupied by $n$ phonons.  
  The advanced mechanism is also phonon-assisted but with split first-order states $\ket{e,1}$ and $\ket{g,1}$. The rightmost panel shows the partial Huang-Rhys factors (HR) and the photoluminescence line shape for $\mathrm{C_2C_B}$ in black and red, respectively. This photoluminescence line shape is obtained with the HSE06 functional and acoustic phonons. Further details are presented in the Methods.
  (b)~The spectral ranges of two optical PSB are shaded in orange and yellow on top of the photoluminescence spectrum at $\SI{168}{meV}$ laser detuning. The bandwidth for the PSB intensities is $\SI{10}{meV}$, identical to the one used for the ZPL intensity. The spectral range of the ZPL (green) is identical to the dashed lines in Fig.~\ref{fig2:PLE}a. We assign the peak around $\SI{270}{meV}$ to another luminescent centre while the peak around $\SI{340}{meV}$ corresponds to the second-order PSB.
  (c)~ZPL and PSB intensities in colors corresponding to~(b). We highlight that the ZPL intensity is identical to Fig.~\ref{fig2:PLE}c. (d)~ZPL intensity in green and flipped photoluminescence in blue. The first- and second-order phonon states are shaded in colors corresponding to the advanced mechanism in~(a). The vertical dashed lines are at detunings of $\SI{160}{}$ and $\SI{190}{meV}$, respectively.
  We note that all photoluminescence (PL) and PLE intensities are background-corrected and normalised (see Supplementary Information~\ref{subsec:SI_BackgroundAndPSB} and~\ref{subsec:SI_Fig4ExtendedDetunings}).}
\end{figure*}

To model the PLE results precisely, we use the relatively more accurate HSE06 functional and study the photoluminescence line shape of Emitter~A in more detail.
The experimental and theoretical line shape disagree at ZPL detunings from $\SI{10}{}$ to $\SI{50}{meV}$, as shown by the inset of Fig.~\ref{fig1:Scatter}b.
While this spectral range was previously associated with another, independent electronic transition~\cite{BommerTwoDefects2019}, we assign this range to a low-energy, acoustic PSB from the \emph{same} electronic transition since its PLE characteristic shows excellent qualitative agreement with the ZPL (see Supplementary Information~\ref{subsec:SI_Fig4ExtendedDetunings}).\par
The experimentally observed acoustic PSB is not reproduced by our \emph{ab initio} calculations, since the latter are carried out in a perfectly planar system with high symmetry that exhibits vanishing coupling to acoustic phonons, i.e. the corresponding partial Huang-Rhys factors are zero. 
This is in contrast to our experiments where the symmetry can be broken by local strain, non-radiative defects close by, non-parallel hBN layers as well as the vicinity of edges, kinks, and grain boundaries. Such a symmetry breaking allows the coupling to acoustic phonons which we observe in several experimental photoluminescence line shapes as asymmetric ZPL lines (see Supplementary Information~\ref{subsec:SI_FurtherGroupISpectra}). 
We highlight that the asymmetric ZPL observed in our experiments has been reported in several works on group~I centres~\cite{RobustTran2016,VoglFabrication2018,FischerSciAdv2021,MendelsonCVD2019,MendelsonCarbon2021,WiggerPhononModelling2019} independent on the generation process.\par

To account for the aforementioned coupling to acoustic phonons, we modify the spectral density shown in Equation~(\ref{eq:SpectralDensityAbInitio}). We include acoustic phonons by adding a Gaussian contribution such that the total spectral density is written as $J(\omega) = J_\mathrm{Sim}(\omega)+J_\mathrm{acoustic}(\omega)$ with
\begin{align*}
J_\mathrm{acoustic}(\omega) = \alpha\omega_c^{-2}\omega\exp(-\omega^2/\omega_c^2),
\end{align*}
where $\alpha$ and $\omega_c$ are found by fitting the photoluminescence spectrum. 
This form of spectral density is commonly used to describe the bulk acoustic phonons in three dimensions resulting from deformation potential coupling observed in semiconductor quantum dots~\cite{Nazir_PolaronMethodReview2016,Kuhn2002}.
We find that the experimental photoluminescence at small ZPL detunings is closely resembled by calculations that include acoustic phonons, showing a very precise theoretical description (see inset of Fig.~\ref{fig1:Scatter}b).
We refer the reader to the Methods section for more details on the theory and the fitting procedure used.\par

With the total spectral density at hand, we calculate the complete PLE spectrum of several defect transitions. 
While the commonly-used generating function approach is limited to the comparison of photoluminescence line shapes, our polaron-based method allows to directly calculate the theoretical ZPL intensity by introducing an external driving field into the system Hamiltonian.
For each laser detuning we calculate the ZPL intensity by sweeping out the experimental detuning range. We highlight that for the obtained ZPL intensities we use the phonons of the ground state, i.e. the identical phonons utilised for the emission line shape. To the best of our knowledge, this is the first time that PLE measurements of $\SI{2}{eV}$ luminescent centres in hBN are replicated by a theoretical model.\par

Fig.~\ref{fig2:PLE}b shows the complete PLE map of $\mathrm{C_2C_B}$ while the maps of $\mathrm{C_2C_N}$ and $\mathrm{V_NC_B}$ are shown in Supplementary Information~\ref{subsec:JakeFitExperiments}. We find excellent agreement of the experimental PLE map with $\mathrm{C_2C_B}$ and $\mathrm{C_2C_N}$, reflecting a very good match of both emission line shape as well as ZPL and PSB intensities. The theoretical PLE of $\mathrm{C_2C_B}$ shows a slightly stronger enhancement at $\SI{341}{meV}$ detuning compared to the experimental data. This may be due to other nonradiative decay channels like shelving states~\cite{BollPhotophysics2020} which are not considered in our model.\par

To compare our PLE experiments with several defect transitions, we study the ZPL intensity which is defined as the area under the ZPL (indicated by vertical lines in Fig.~\ref{fig2:PLE}a and~b). Fig.~\ref{fig2:PLE}c shows excellent agreement of the experimental ZPL intensity with $\mathrm{C_2C_B}$ and $\mathrm{C_2C_N}$ while $\mathrm{V_NC_B}$ shows poorer agreement, thus allowing us to exclude the latter defect for Emitter~A.\par

All in all, our excellent agreement of the complete experimental PLE with our polaron-based calculations for $\mathrm{C_2C_B}$ and $\mathrm{C_2C_N}$ is outstanding compared to previous work~\cite{GrossoPLE2020}. This finding infers that, among the studied defect transitions, $\mathrm{C_2C_B}$ and $\mathrm{C_2C_N}$ are the most likely microscopic origins for Emitter~A. 
We highlight that the generation mechanism of Emitter~A, outlined in Ref.~\cite{FischerSciAdv2021}, also holds for $\mathrm{C_2C_B}$ and $\mathrm{C_2C_N}$, since both defects can be formed by merging of $\mathrm{C_N}$ and $\mathrm{C_B}$ during thermal annealing.\par

Previously, experimental photoluminescence line shapes were compared with vacancies~\cite{TranNanoTech2016,LiNonmagnetic2017}, oxygen-based defects~\cite{XuPlasmaSidney2018} and carbon-based defects~\cite{MendelsonCarbon2021,FischerSciAdv2021,JaraC2CN2021}.
Although Emitter~A shows good agreement with $\mathrm{C_2C_B}$ and $\mathrm{C_2C_N}$, 
other luminescent centres, also generated by our process, show agreement with different defects. 
One example is Emitter~C (see Fig.~\ref{fig1:Scatter}d) showing a small Euclidean distance for $\mathrm{V_NC_B}$, i.e. a good agreement with this defect.
Furthermore, our irradiation-based process also generates luminescent centres with line shapes different to group~I centres that show small Euclidean distances to defects different to $\mathrm{C_2C_B}$, $\mathrm{C_2C_N}$, and $\mathrm{V_NC_B}$ (see Supplementary Information~\ref{subsec:NoGroupISpectra}). 
We are convinced that our process generates several types of defects that appear as $\SI{2}{eV}$ luminescent centres in our experiments. Therefore, comparing PLE maps of \emph{individual} luminescent centres with our polaron method helps to identify their microscopic origins.\par

\textbf{Excitation mechanism}\\
The microscopic origin of twelve group~I centres has been narrowed down to three defect transitions~(Fig.~\ref{fig1:Scatter}). 
These three transitions have relatively large Huang-Rhys factors around $\SI{170}{meV}$ and thus can be modelled as two-level systems with discrete vibronic energy levels (see Supplementary Information~\ref{subsec:SI_Fig3TheoryDetails}).
To verify the vibronic structure of these energy levels, which are directly related to the excitation mechanism, PLE is a strong tool since it probes off-resonant transitions in both ground and excited states.\par

The excitation mechanism of $\SI{2}{eV}$ luminescent centres in hBN does not involve interband transitions since the photon energy of the excitation laser ($\sim \SI{2.5}{eV}$) is much smaller than the band gap $E_g\sim~\SI{6}{eV}$~\cite{CassaboisBandgaphBN2016}. Supported by charge transfer experiments~\cite{Xu_ChargeTransfer2020}, the excitation mechanism most likely involves only electronic states inside the band gap, so-called deep levels.
This makes the energy of the driving laser very important since only few, discrete electronic states are available - in contrast to a continuum of states for interband transitions in semiconductors.\par

For $\SI{2}{eV}$ luminescent centres in hBN, one possible configuration of these deep levels is the empiric mechanism illustrated in Fig.~\ref{fig3:ExcitationMechanism}a. 
Here, the driving laser excites an electron from a deep level below the Fermi level $E_F$ to another level lying far below the conduction band. From here, the electron relaxes to the excited state $\ket{e}$ by dissipating energy~$\Delta_e$. Then, a photon belonging to the ZPL is emitted by a transition to the ground state $\ket{g}$, followed by another relaxation with energy~$\Delta_g$.\par

Recent PLE experiments~\cite{GrossoPLE2020,MaleinPLE2021} observed an enhanced ZPL emission at detunings around $\SI{170}{meV}$. 
While the authors assigned these results to a phonon-assisted process, the empiric mechanism can in principle also be the underlying excitation mechanism with $\Delta_e+\Delta_g~\sim~\SI{170}{meV}$. On the contrary, our PLE measurements show enhanced ZPL intensities both at detunings around~$\SI{170}{}$ \emph{and}~$\SI{340}{meV}$ (see Fig.~\ref{fig2:PLE}c). 
The empiric mechanism could be expanded by another electronic level, but the simplest explanation of our PLE results is the phonon-assisted excitation mechanism.\par

\textbf{Phonon-assisted mechanism}\\
As shown above, our PLE measurements can not be explained by the empiric mechanism. Therefore, we introduce the phonon-assisted excitation mechanism, also called Huang-Rhys model (see Fig.~\ref{fig3:ExcitationMechanism}a) which was proposed in previous works on hBN~\cite{JungwirthExcitationScheme2017,GrossoPLE2020}. 
Here, two electronic states couple to one phonon mode with energy $\hbar \Omega$, resulting in discrete vibronic energy levels $\ket{e,n}$ and $\ket{g,m}$. The driving laser excites a high laying vibronic state $\ket{e,n}$ in the excited electronic manifold which rapidly relaxes to its lowest state $\ket{e,0}$. 
A photon is then emitted through an electronic transition to the ground-state manifold, producing ZPL and PSB emission.\par

By varying the photon energy of the driving laser, we can resonantly excite different vibronic states $\ket{e,n}$ in the excited-state manifold (see Fig.~\ref{fig3:ExcitationMechanism}a). 
In particular, the ZPL and PSB emission are enhanced at equally-spaced laser detunings of $\hbar \Omega$ and $2 \hbar \Omega$, also called one- and two-phonon detuning. We observe this in our PLE experiments by studying the ZPL intensity (see Fig.~\ref{fig2:PLE}c) but also by studying two PSB intensities, as outlined below.\par

To study PSB emissions as a function of laser detuning, we define the PSB intensity as the area under the PSB, indicated by shaded areas in Fig.~\ref{fig3:ExcitationMechanism}b. We find an excellent qualitative agreement of the ZPL intensity with the two PSB intensities that are taken at ZPL detunings of around $\SI{160}{}$ and $\SI{190}{meV}$ (see Fig.~\ref{fig3:ExcitationMechanism}c).
This finding shows that the ZPL and the two optical PSB have the same excitation mechanism and thus can be used to investigate the excitation mechanism of Emitter~A. To the best of our knowledge, this is the first time that ZPL and PSB intensities of $\SI{2}{eV}$ luminescent centres in hBN are directly compared with each other.\par

The ZPL and PSB intensities are enhanced at equally-spaced detunings of $\sim \SI{170}{}$ and $\sim \SI{340}{meV}$, as shown in Fig.~\ref{fig3:ExcitationMechanism}c. This experimental finding agrees excellently well for $\hbar \Omega \sim \SI{170}{meV}$ with the phonon-assisted mechanism where ZPL and PSB emission are enhanced at detunings of $\hbar \Omega$ and $2 \hbar \Omega$.
At detunings between the two aforementioned energies, we obtain small ZPL and PSB intensities because the excitation laser energy is between the two states $\ket{e,1}$ and $\ket{e,2}$.
Furthermore, the photoluminescence (Fig.~\ref{fig3:ExcitationMechanism}b) shows first- and second-order PSB at detunings around $\SI{170}{}$ and $\SI{340}{meV}$ corresponding to optical transitions from $\ket{e,0}$ to $\ket{g,1}$ and $\ket{g,2}$, respectively. Therefore, our experiments allow us to discard the empiric mechanism for Emitter~A and confirm indications of a phonon-assisted mechanism for group~I centres in hBN.
This might pave the way towards a universal excitation mechanism, which could also explain other experimental findings such as high-temperature photoluminescence~\cite{Kianinia800Kelvin2017} and photophysics~\cite{BollPhotophysics2020,TranPRA2016}.\par

In the phonon-assisted excitation mechanism (see Fig.~\ref{fig3:ExcitationMechanism}a), the ZPL emission via $\ket{e,2}$ (two-phonon detuning) and the PSB emission via $\ket{e,1}$ (one-phonon detuning) are both processes involving two phonons, and thus should show comparable intensities (details in Methods). Indeed, our experimental ZPL intensity at two-phonon detuning shows similar strength as both PSB intensities at one-phonon detuning, since both processes involve two phonons (see Supplementary Information~\ref{subsec:SI_Fig4ExtendedDetunings}). Furthermore, the ZPL intensity via $\ket{e,1}$ involves only one phonon and thus is stronger than both aforementioned two-phonon processes. These results further support the phonon-assisted excitation mechanisms of Emitter~A.\par

\textbf{Detailed phonon coupling}\\
The experimental photoluminescence of Emitter~A shows a first-order PSB at detunings around $\SI{170}{meV}$ which consists of two distinct peaks at $\SI{160}{}$ and $\SI{190}{meV}$, shown by dashed lines in Fig.~\ref{fig3:ExcitationMechanism}d. Furthermore, we find that the ZPL intensity also shows a strong peak at detunings around $\SI{170}{meV}$ with two shallower peaks at $\SI{160}{}$ and $\SI{190}{meV}$. This reveals that both first-order states~$\ket{e,1}$ and~$\ket{g,1}$ are split into two distinct levels, reflected in the advanced excitation mechanism (see Fig.~\ref{fig3:ExcitationMechanism}a).
Here, the second-order phonon replicas~$\ket{e,2}$ and~$\ket{g,2}$ are shown as continua since our experiments do not show distinct peaks at two-phonon detuning (see Fig.~\ref{fig3:ExcitationMechanism}d).
The calculated HR factors of $\mathrm{C_2C_B}$ (Fig.~\ref{fig3:ExcitationMechanism}a) as well as split first-order PSB of several group~I centres (see Supplementary Information~\ref{subsec:SI_FurtherGroupISpectra}) support that several phonon modes are involved in the excitation mechanism of group~I centres in hBN.
We note that previous works have assigned a split $\ket{g,1}$ to longitudinal and transverse optical bulk phonons~\cite{WiggerCoherent2022,FeldmanCrosscorrelation2019,KhatriPhononAssignment2019}. In our work, we provide a more accurate description since we calculate the phonon modes coupled to the defect in the lattice, thereby taking into account both localised and delocalised phonon modes.\par

In Fig.~\ref{fig3:ExcitationMechanism}d we compare the ZPL intensity with the flipped photoluminescence line shape. For the $\SI{160}{}$ and $\SI{190}{meV}$ peaks we observe a blueshift and a redshift between the photoluminescence and ZPL intensity, respectively. This could point towards different phonon coupling between ground and excited states. While the redshift of the $\SI{190}{meV}$ peak is in agreement with previous work~\cite{MaleinPLE2021}, the blueshift for the $\SI{160}{meV}$ peak has not been observed. Both spectral shifts (around $\SI{10}{meV}$) are, however, on the same order of magnitude as the spectral resolution of both PLE setups (around $\SI{5}{meV}$) and thus further PLE experiments with higher spectral resolution are needed to confirm the presence of fine differences in the phonon coupling between ground and excited states.\par


\section*{Conclusion}
In summary, we have studied the photoluminescence of luminescent centres in hBN with a focus on group~I centres showing pronounced PSB around $\SI{170}{meV}$ and a ZPL energy around $\SI{2}{eV}$. By combining \emph{ab initio} methods with the non-perturbative polaron method, we calculated the emission line shapes of 26 different defect transitions.
Studying both the theoretical ZPL energy and the Euclidean distance between experiment and theory showed that $\mathrm{C_2C_B}$, $\mathrm{C_2C_N}$, and $\mathrm{V_NC_B}$ are the most likely defect candidates for twelve experimental line shapes.
Our method of using the Euclidean distance represents a new tool to narrow down the number of possible defects responsible for luminescent centres in insulators as well as to make accurate predictions of their optical properties when integrated in photonic structures.\par

Our method of analysis represents a complement to the common methodology, which compares theoretical values of the total Huang-Rhys or Debye-Waller factor with values extracted from the experiment. Indeed, in this work, variations in the intensity $S(\Delta_i)$ due to the finer structure in the PSB are taken into account through the variance $\sigma$.
Capturing these fine details in the PSB is crucial to construct accurate models of the electron phonon coupling dynamics. Indeed, large variations in the spectral density caused by spectrally sharp phonon modes can induce non-trivial phonon-electron-photon correlations~\cite{kamper2022signatures}.\par

The full potential of our new approach, combining \emph{ab initio} calculations with the polaron method, comes into light when compared to PLE measurements of our group~I centers. In contrast to the generating function approach, our polaron model allows introducing an external driving field and thereby enables us to calculate theoretical PLE maps.
By focusing on the most likely defect candidates and adding acoustic phonons to the spectral density, we found excellent agreement of the experimental PLE of one group~I centre with $\mathrm{C_2C_B}$ and $\mathrm{C_2C_N}$ while we could exclude 24 other defect transitions. Our excellent agreement of experimental PLE with \emph{ab initio} calculations is outstanding compared to previous work~\cite{GrossoPLE2020}.\par

In our PLE experiments, we observed enhanced ZPL and PSB emission at one- and two-phonon detunings, the latter for the first time to the best of our knowledge. 
By resolving first- and second-order phonon transitions, we confirmed indications of a phonon-assisted excitation mechanism with a phonon energy around $\SI{170}{meV}$. 
In particular, it is very unlikely that other electronic states are equally-spaced by $\SI{170}{meV}$ for both ground and excited states.
Moreover, such additional electronic states are not predicted by \emph{ab initio} calculations for the most likely defect transitions. Our findings are supported by split first-order PSB of several group~I centres as well as by a decent agreement of our excitation mechanism with \emph{ab initio} calculations of $\mathrm{C_2C_B}$. 
The methodology of combining PLE measurements and the polaron method with input from \emph{ab initio} methods can be applied to identify the microscopic origin of other luminescent centres or single-photon emitters in hBN and other materials.\par


We are convinced that the presented comparison of experimental PLE maps with advanced models, able to combine atomistic calculations with open quantum system theory, provides the most accurate description of $\SI{2}{eV}$ luminescent centres in hBN and their excitation mechanism.
For luminescent centres showing excellent agreement with $\mathrm{C_2C_B}$ and $\mathrm{C_2C_N}$, lifetime measurements can give further insight since the theoretical values differ significantly~\cite{LiSmartPing2022}.
To prove the quantum nature of the studied luminescent centres, auto-correlation ($g^{(2)}$ function) and cross-correlation measurements~\cite{FeldmanCrosscorrelation2019} as well as polarisation-dependent measurements~\cite{JungwirthExcitationScheme2017,Ates2023Polarization} are required.
Furthermore, bleaching~\cite{MendelsonCVD2019,Martinez_Bleaching2016} and blinking~\cite{FischerSciAdv2021,TranPRA2016,LiSiteSelective2021} of $\SI{2}{eV}$ luminescent centres are unresolved challenges since only few centres were stable over minutes~\cite{TranNanoTech2016,Kianinia800Kelvin2017,FeldmanCrosscorrelation2019,XuNanoindentation2021,FischerSciAdv2021} or even months~\cite{VoglFabrication2018}.
All in all, further experimental and theoretical work is needed to both unambiguously identify the nature of group~I centres and to make $\SI{2}{eV}$ luminescent centres in hBN feasible for applications in quantum technologies.\par

As a last observation, we would like to point out that the ubiquitously used $\SI{532}{nm}$ laser efficiently excites luminescent centres with ZPL energies around $\SI{2.15}{eV}$ because the laser energy of $\SI{2.33}{eV}$ matches the one-phonon detuning of around $\SI{170}{meV}$. 
This fortunate coincidence of the energy of an ubiquitously used laser with the one-phonon detuning has been a gift for the research field and explains why the majority of works on hBN report $\SI{2}{eV}$ luminescent centres.\par

\section*{Methods}
\textbf{Scatter plot}\\
The mean and variance shown in Fig.~\ref{fig1:Scatter}c and Supplementary Information~\ref{subsec:DetailedScatterPlots} are calculated for ZPL detunings from $\SI{-440}{}$ to $\SI{10}{meV}$. We note that the only fitting parameter is the ZPL line width while the phonon broadening is kept constant (see Supplementary Information~\ref{subsec:JakeFitExperiments}).\par
The scatter plots reflect the agreement of the partial Huang-Rhys factors of each defect transition with the experimental line shape. The discussed scatter plots are similar to a recent theoretical study~\cite{TheoryLinderalv2021} focusing on the combined defect and PSB emission spectrum (i.e. accessing the vibrational fine structure).\par
We note that in our scatter plots we focus on the mean and variance while higher-order variances need to be considered for perfect agreement between theory and experiment.\par

\textbf{\emph{Ab initio} calculations}\\
All the \emph{ab initio} calculations for the ground states, excited states and normal modes were performed within the GPAW electronic structure code~\cite{enkovaara2010electronic} using a plane-wave basis set with $\SI{800}{eV}$ plane wave cut-off and a $\Gamma$-point sampling of the Brillouin zone.
All the defects were represented in a 7x7x1 supercell (monolayer) and allowed to fully relax until the maximum force was below $\SI{0.01}{eV\per\angstrom}$. A vacuum of $\SI{15}{\angstrom}$ was used in the vertical direction.\par
The PBE exchange correlation (xc-)functional~\cite{perdew1996generalized} was used for all calculations. In addition, for $\mathrm{C_2C_B}$, $\mathrm{C_2C_N}$, and $\mathrm{V_NC_B}$ the relatively more accurate HSE06 functional was used for the calculation of the ZPL energies.
The excited states were calculated using the direct optimisation-maximum overlap matrix (DO-MOM) method~\cite{levi2020variational} with a maximum step length, $\rho_\mathrm{max}$, for the quasi-Newton search direction of $0.2$. Compared to the standard $\Delta$-SCF approach, the DO-MOM method yields improved convergence and avoids variational collapses during the SCF optimisation~\cite{levi2020variational}. 
We highlight that the $\mathrm{C_2C_N}$ described here is the transition studied in Ref.~\cite{LiSmartPing2022}.
For further details on the theory and calculation details we refer to our previous work~\cite{BertoldoQPOD2022}.\par
In all calculations, we considered in-plane defect structures and verified their dynamical stability by the absence of imaginary frequencies in the $\Gamma$-point phonon spectrum. Only for the excited state of $\mathrm{V_NC_B}$, i.e. $(2)^{3}B_\text{1}$, does the carbon atom of the defect relax out of plane. The out-of-plane configuration is stable compared to the planar configuration by $\SI{0.23}{eV}$ for monolayer hBN. This energy difference is only $\SI{0.036}{eV}$ in a trilayer hBN with a $\mathrm{V_NC_B}$ defect embedded in the central hBN layer (of the trilayer hBN)~\cite{FischerSciAdv2021}.

\textbf{Optical spectroscopy}\\
The optical characterisation is described in detail in Supplementary Information~\ref{subsec:SI_Optical characterisation}. For PLE measurements at low temperature, we used a supercontinuum white light laser 
in combination with a tunable laser line filter 
resulting in a PLE resolution of $\sim \SI{5}{meV}$.\par

\textbf{Open quantum system}\\
To model experimental PLE maps we use information about the phonon modes extracted from the \emph{ab initio} calculations to construct an open system model for the defect transition. 
The emitter is modelled as a two level system with ground and excited states $\ket{g}$ and $\ket{e}$ respectively, with transition energy $\hbar\omega_e$. 
The emitter is driven by a continuous-wave laser with angular frequency $\omega_\mathrm{L}$, and interacts both with vibrational and electromagnetic environments. The form of this interaction is given in the Supplementary Information~\ref{subsec:JakeModel}. 
To account for strong electron-phonon coupling and resultant PSB observed for luminescence centres, we make use of the polaron method~\cite{mccutcheon2010quantum,Nazir_PolaronMethodReview2016}. 
This approach dresses the system energy levels with modes of the vibrational environment using a unitary transformation, providing an optimised basis in which to do perturbation theory.
We then derive a Born-Markov master equation in the transformed frame which is non-perturbative in the original electron phonon interactions. 
This allows us to describe the dynamics~\cite{mccutcheon2010quantum} and, crucially, the PSB of the optical transition in question~\cite{iles2017limits,iles2017phonon}.
For further details of the model described above, we refer the reader to the Supplementary Information~\ref{subsec:JakeFitExperiments} and~\ref{subsec:JakeModel}.\par
We note that the expression of $J_\mathrm{acoustic}(\omega)$ uses a different definition of the spectral density compared to Ref.~\cite{Nazir_PolaronMethodReview2016} which can be obtained by multiplying $J_\mathrm{acoustic}$ with $\omega^2$ and $\alpha$ with $\omega_c^{-2}$. The physical description remains unchanged, i.e. in both cases a three-dimensional acoustic phonon environment is used.\par

\textbf{Excitation mechanism}\\
In Fig.~\ref{fig3:ExcitationMechanism}a, the band gap is given by $E_g \sim \SI{6}{eV}$~\cite{CassaboisBandgaphBN2016} and the Fermi level is $E_F$. Within the empiric mechanism, the two relaxation processes $\Delta_e$ and $\Delta_g$ are associated with $\ket{e}$ and $\ket{g}$, respectively.
In the phonon-assisted and advanced excitation mechanisms, we do not show the Fermi level for clarity. Furthermore, the excited states are labelled with $\ket{e,n}$ corresponding to the electron in the excited state and its environment occupied by $n$ phonons. Similarly, $\ket{g,m}$ describes the electron in the ground state with $m$ phonons in its environment.\par

\textbf{One- and two-phonon processes}\\
To identify one- and two-phonon processes, we use the electronic states $\ket{e,n}$ and $\ket{g,m}$ of the phonon-assisted excitation mechanism presented in Fig.~\ref{fig3:ExcitationMechanism}a.\par
At a laser detuning of $\sim \SI{170}{meV}$, we excite the electronic state $\ket{e,1}$ that relaxes to $\ket{e,0}$ by the emission of one phonon. From $\ket{e,0}$, the ZPL emission is realised without any further phonon emission, while the PSB process involves another phonon that is emitted by the relaxation from $\ket{g,1}$ to $\ket{g,0}$. Therefore, the ZPL is a one-phonon process while the PSB is a two-phonon process. The above described two-phonon process holds for both optical PSB at~$\SI{160}{}$ and~$\SI{190}{meV}$ as well as the acoustic PSB at $\sim \SI{10}{meV}$.\par
At laser detunings of $\sim \SI{340}{meV}$ we excite the state $\ket{e,2}$, introduced in the phonon-assisted mechanism in Fig.~\ref{fig3:ExcitationMechanism}a. This state relaxes by the emission of two phonons to $\ket{e,0}$ from where a ZPL emission is realised without any further phonon emission. Thus, this process is also a two-phonon process like the optical PSB at $\sim \SI{170}{meV}$ detuning.\par


\subsection*{Acknowledgements}
\textbf{Funding:} This work was funded by the Danish National Research Foundation through the Center for Nanostructured Graphene (project number DNRF103) and through NanoPhoton - Center for Nanophotonics (project number DNRF147). M.F. and N.S. acknowledge support from the VILLUM FONDEN (project number 00028233). Parts of our optical setup are financed through the IDUN Center of Excellence founded by the Danish National Research Foundation (project number DNRF122) and VILLUM FONDEN (project number 9301). N.S. and M.W. acknowledge support from the Independent Research Fund Denmark, Natural Sciences (project number 0135-00403B). S.X. acknowledges support from the Independent Research Fund Denmark, Technology and Production Sciences (project number 9041-00333B) and Natural Sciences (project number 2032-00351B). D.I.M. and S.C. acknowlege support from the Independent Reseach Fund Denmark, Sapere Aude grant (project number 8049-00095B).
S.A. and K.S.T. acknowledge funding from the European Research Council (ERC) under the European Union’s Horizon 2020 research and innovation program Grant No. 773122 (LIMA) and the Novo Nordisk Foundation Challenge Programme 2021: Smart nanomaterials for applications in life-science, BIOMAG Grant NNF21OC0066526. K.S.T. is a Villum Investigator supported by VILLUM FONDEN (grant no. 37789).
A.H., C.K., and A.W.H. gratefully acknowledge the German Science Foundation (DFG) for financial support via the clusters of excellence MCQST (EXS-2111) and e-conversion (EXS-2089), as well as the state of Bavaria via the One Munich Strategy and the Munich Quantum Valley, and TUM-IGSSE for project BrightQuanDTUM.\par

\subsection*{Author contributions}
M.F. and A.H. carried out the experiments. M.F. and N.S. analysed and interpreted the experimental data. A.S. designed and performed the \emph{ab initio} calculations. K.S.T. supervised the \emph{ab initio} calculations. J.I.-S. developed the open quantum systems model, and fitted the theoretical curves to the experimental data. M.F., A.S., J.I.-S., A.H., K.S.T., A.W.H. and N.S. discussed the experimental results and compared them with the theoretical predictions. A.S. and J.I.-S. equally contributed to the theoretical part of this manuscript. All authors contributed to the writing of the manuscript.

\subsection*{Note}
During the preparation of this manuscript we became aware about a thorough study of defects in hBN, focusing on tailoring the ZPL energy~\cite{VoglTailoringWavelength2022}.

\subsection*{Competing interests}
The authors declare that they have no competing interests.

\onecolumngrid
\newpage
\section*{Supplementary Materials\\
Combining experiments on luminescent centres in hexagonal boron nitride with the polaron model and ab initio methods towards the identification of their microscopic origin}

\section{Further group~I spectra}\label{subsec:SI_FurtherGroupISpectra}
\begin{figure}[h]
  \includegraphics[width=0.8\columnwidth]{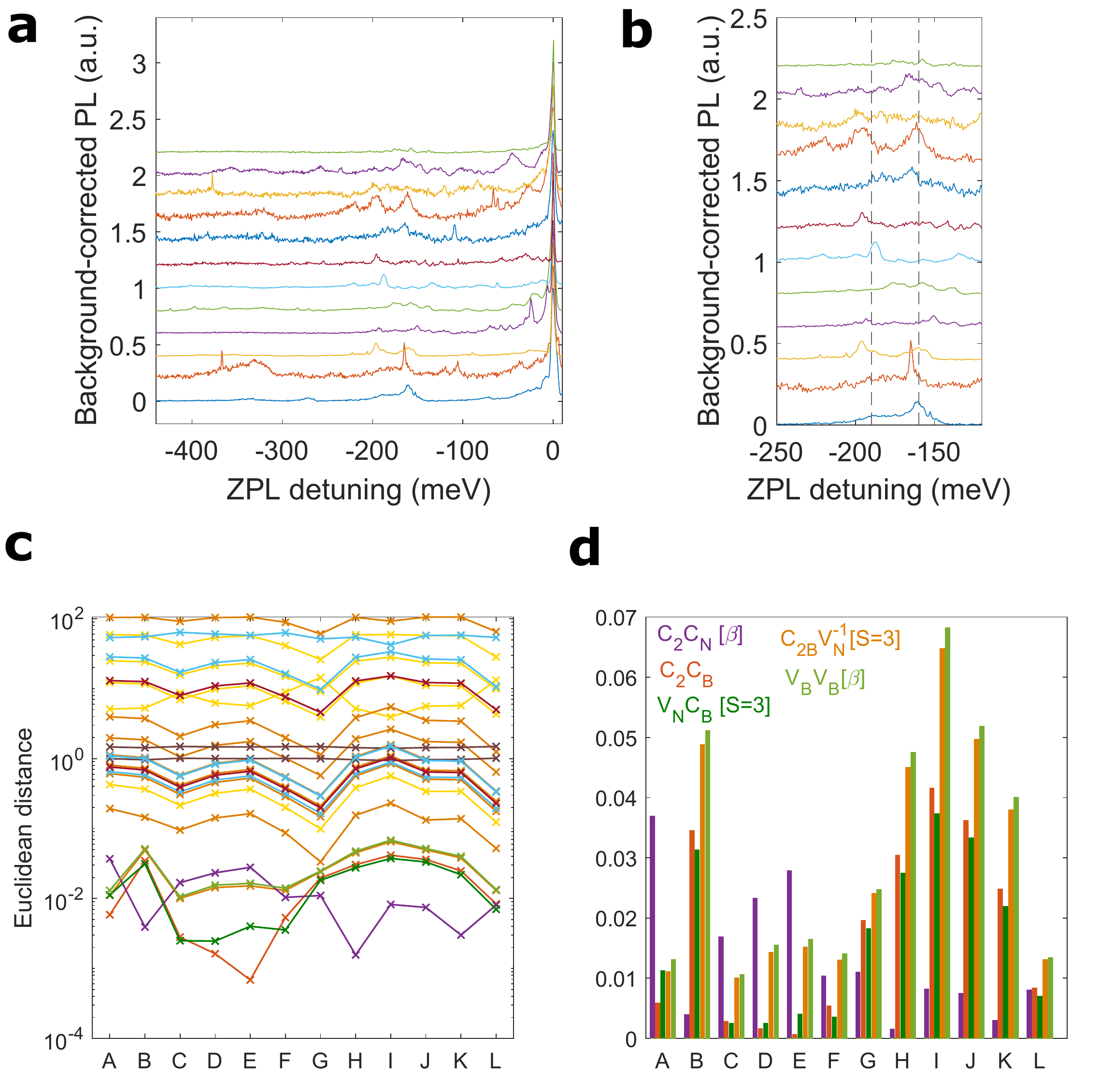}
  \caption{\label{fig1_SI_12GroupI}\textbf{Low-temperature photoluminescence of luminescent centres belonging to group~I.} (a)~Photoluminescence line shapes at $T = \SI{10}{K}$ under $\SI{2.37}{eV}$ excitation, shifted vertically for clarity. The spectra from bottom to top correspond to Emitter~A to Emitter~L, given by the x axis in~(c). (b)~Optical PSB of the luminescent centres shown in (a) with corresponding colors, shifted vertically for clarity. The vertical dashed lines are at detunings of $\SI{160}{}$ and~$\SI{190}{meV}$, respectively.
  (c)~Euclidean distance for Emitter~A up to Emitter~L, with corresponding spectra shown in~(a) from bottom to top. (d)~Histograms of Emitter~A to Emitter~L. The colors correspond to the defect transitions given by the legend.  
  All line shapes are background corrected as outlined in Supplementary Information~\ref{subsec:SI_BackgroundAndPSB} and more details are given in the main text.}
\end{figure}

\begin{table}[h]
    \centering
    \begin{tabular}{c|c|c|c|c|c|c|c|c|c|c|c}
         A  &   B   &   C   &   D   &   E   &   F   &   G   &   H   &   I   &   J   &   K   &   L\\\hline
    2.1528  &2.1785 &2.0122 &2.2500 &2.1797 &2.2045 &2.2133 &2.1652 &2.2058 &2.0819 &2.1703 &2.1984
    \end{tabular}
    \caption{Zero-phonon line energies (in electron-volt) of Emitter~A to~L.}
\end{table}

\newpage
\section{Details on \emph{ab initio} calculations}\label{subsec:SI_Fig3TheoryDetails}

\textbf{Nomenclature of defect transitions}\\
In the main text, we suppress the notations for the spin majority channel $\alpha$ as well as non-triplet spin states. For $\mathrm{C_2C_B}$, this results in
\begin{align}
    \mathrm{C_2C_B[S=2,\alpha]} \equiv \mathrm{C_2C_B}~,
\end{align}
where $\equiv$ stands for an equal representation. Another example is $\mathrm{C_2C_N[S=2, \beta]} \equiv \mathrm{C_2C_N[\beta]}$.

\begin{table*}[h]
    \centering
    \begin{tabular}{ll|cccccc}
    \hline \hline
    &&\multicolumn{3}{c}{$\alpha$}&\multicolumn{3}{c}{$\beta$}\\
    Defect                  &Charge and spin state                  &ZPL(eV)    &$\Delta$Q  &HR     &ZPL(eV)    &$\Delta$Q  &HR\\ \hline
    $\mathrm{C_2C_N-V_N}$   &$\mathrm{C_2C_N-V_N}$ [S=3](E=2.15eV)  &0.2        &1.3        &7.45   &0.96       &1.76       &19.7\\
                            &$\mathrm{C_2C_N-V_N}$ [S=0](E=0.0eV)   &1.96   	&0.8        &5.72   &--         &--         &--\\		
                            &$\mathrm{C_2C_N-V_N^{+1}}$ [S=2] 	    &--         &--         &--   	&0.72	    &1.42	    &15.7\\
                            &$\mathrm{C_2C_N-V_N^{-1}}$ [S=2]       &--         &--         &--     &1.96       &1.3        &9.1\\
    $\mathrm{C_{2N}V_N}$    &$\mathrm{C_{2N}V_N}$ [S=2]             &2.08	    &1.083	    &7.16	&--   	    &--   	    &--\\
                &$\mathrm{C_{2N}V_N^{+1}}$ [S=0](E=0.0eV)           &2.2	    &0.83	    &4.14   &--         &--         &--\\
                &$\mathrm{C_{2N}V_N^{-1}}$ [S=3](E=1.43eV)          &1.7        &0.72       &3.82	&1.07       &0.41	    &1.29\\
                &$\mathrm{C_{2N}V_N^{-1}}$ [S=0](E=0.0eV)           &1.47	    &1.31	    &9.05   &--         &--         &--\\
    $\mathrm{C_{2B}V_N}$    &$\mathrm{C_{2B}V_N}$ [S=2]             &1.41	    &1.08	    &5.67	&1.1	    &0.8	    &5.07\\
                &$\mathrm{C_{2B}V_N^{-1}}$ [S=0]                    &0.77	    &0.61	    &3.2	&--         &--         &--\\
                &$\mathrm{C_{2B}V_N^{-1}}$ [S=3](E=0.64eV)          &0.9        &0.29	    &0.635	&1.17	    &0.8	    &4.99\\
                &$\mathrm{C_{2B}V_N^{+1}}$ [S=0](E=0.0eV)           &0.7        &0.9	    &5.56	&--         &--         &--\\
                &$\mathrm{C_{2B}V_N^{+1}}$ [S=3](E=0.64eV)          &1.12       &1.39       &9.91   &0.74       &1.19       &10.8\\
    $\mathrm{C_2C_B-V_B}$&$\mathrm{C_2C_B-V_B}$ [S=3](E=0.0eV)	    &1.39	    &3.05	    &45	    &--   	    &--       	&--\\
                         &$\mathrm{C_2C_B-V_B^{-1}}$ [S=2]          &0.74	   &13.52	   &36.8   &--	       &--         &--\\
    $\mathrm{V_BV_B}$ &$\mathrm{V_BV_B}$ [S=3]                      &--         &--   	    &--     &2.7   	    &0.41   	&0.3\\
    $\mathrm{C_BC_NV_N}$ &$\mathrm{C_BC_NV_N^{+1}}$ [S=3](E=3.92eV) &1.37       &0.92   	&10.5	&--         &--         &--\\
                         &$\mathrm{C_BC_NV_N^{+1}}$ [S=0](E=0.0eV)  &2.95       &1.19   	   &7.45   &--         &--         &--\\
    $\mathrm{C_2C_N}$&$\mathrm{C_2C_N}$ [S=2]                       &--         &--         &--     &1.51       &0.25       &1.2\\
    $\mathrm{C_2C_B}$&$\mathrm{C_2C_B}$ [S=2]                       &1.16       &0.24       &1.1    &--         &--         &--\\
    $\mathrm{V_NC_B}$&$\mathrm{V_NC_B}$ [S=3]                       &1.39       &0.53       &1.58   &--         &--         &--\\
    $\mathrm{C_2C_N}$, HSE06&$\mathrm{C_2C_N}$ [S=2]                &--         &--         &--     &1.67       &0.26       &1.26\\
    $\mathrm{C_2C_B}$, HSE06&$\mathrm{C_2C_B}$ [S=2]                &1.36       &0.26       &1.2    &--         &--         &--\\
    $\mathrm{V_NC_B}$, HSE06&$\mathrm{V_NC_B}$ [S=3]                &1.75       &0.53       &1.50   &--         &--         &--\\
    \hline \hline
    \end{tabular}
    \caption{\textbf{Zero-phonon lines (ZPL), momentum displacements ($\Delta$Q) and Huang-Rhys factors (HR) of all studied defects.} Here, $\alpha$ refers to the majority spin channel and $\beta$ to the minority spin channel. The energy $E$ in the second column gives the energy difference between the singlet and triplet states. It is only given for defects that show both a singlet and a triplet state. 
    The PBE functional is used for all calculations except for the last three rows where the HSE06 functional was used.
    }
\end{table*}

\newpage
\begin{figure*}[h]
    \centering
    \includegraphics[width=\textwidth]{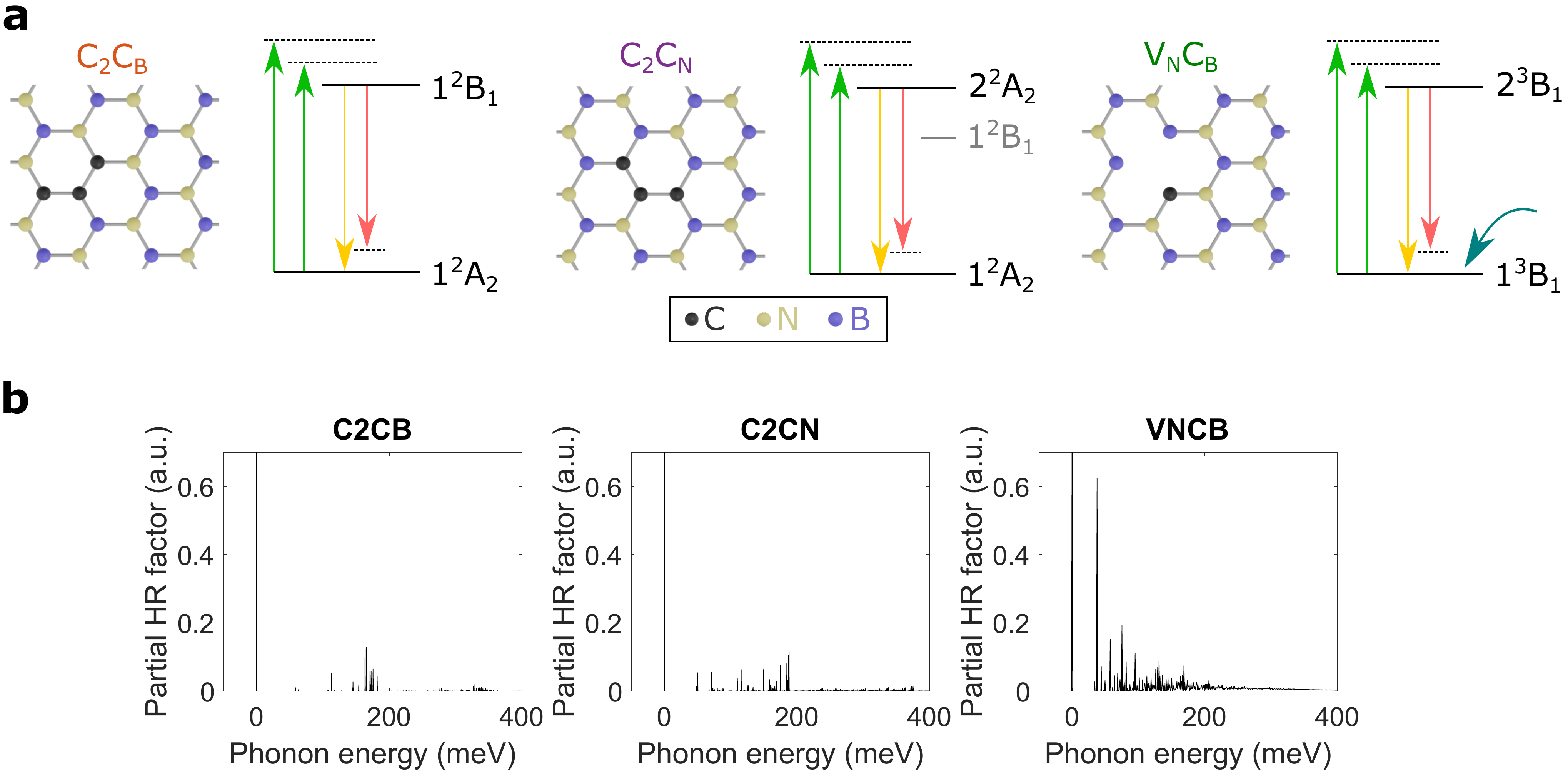}
    \caption{\textbf{Schematics, electronic levels and partial Huang-Rhys factors for $\mathbf{C_2C_B}$, $\mathbf{C_2C_N}[\beta]$, and $\mathbf{V_NC_B}[S=3]$.} 
    (a)~Schematics and electronic levels of $\mathrm{C_2C_B}$, $\mathrm{C_2C_N}[\beta]$, and $\mathrm{V_NC_B}[S=3]$ where the latter is taken from Ref.~\cite{Sajid_PRB_VNCB2018}.
    (b)~The partial Huang-Rhys factors for $\mathrm{C_2C_B}$, $\mathrm{C_2C_N}[\beta]$, and $\mathrm{V_NC_B}[S=3]$ are shown from left to right. We highlight that the HSE06 functional was used for all three defects.}
\end{figure*}

\newpage
\section{Fitting and comparison with experiment}\label{subsec:JakeFitExperiments}

In this section we outline the fitting procedure used to compare the polaron method to experiment.
The fitting is done in two parts, in the first we extract the width of ZPL, the second fits the inhomogeneous broadening about the ZPL associated to acoustic phonons.
To do both of these fits, we consider the emission lineshape driven by a excitation laser with detuning $\tilde{\Delta} = 168$~meV. 
We also assume that the phonons decay on a much faster timescale than light emission occurs, which allows us to ignore the excitation effects during the fitting procedure. 
This allows us to disregard the coherent driving term $\Omega_\mathrm{R}\sigma_x/2$ and the phonon dissipator $\mathcal{K}_\mathrm{PH}$ in Eq.~\ref{eq:meq}, and assume that the emitter is initially in its excited state.

\subsection{Fitting the zero-phonon line}
To fit the ZPL, we restrict the experimental data to a frequency window $\pm 0.008$~eV around the measured ZPL. This allows us to ignore the phonon sideband contribution, fitting only the ZPL with only function $S_\mathrm{opt}(\omega)$. For an emitter initially in its excited state, the ZPL spectrum reduces to a simple Lorentzian lineshape:
\begin{equation}
 S_\mathrm{opt}(\Delta\omega) \approx \frac{B^2}{2\Gamma}\frac{\Gamma+ \gamma}{(\Gamma + \gamma)^2
+\Delta\omega^2},
\end{equation}
where $\Delta\omega$ is the detuning from the emitter frequency. 
After first normalising the data and the above expression to the peak of the ZPL, we do a least min-squared fit. Note that we can fit only a single rate, since $\Gamma$ and $\gamma$ combine to give a single width of the Lorentzian, thus we set $\gamma=0$ without loss of generality. We find an optimal $\Gamma = 3.15$~meV with residual of $r= 0.098$ for Emitter~A.

\subsection{Describing electron-phonon sidebands in photoluminescence}
As discussed in the main tex, we divide the phonon spectral density into two components $J_\mathrm{Ph}(\omega) = J_\mathrm{FP}(\omega) + J_\mathrm{A}(\omega)$.
The first contribution, $J_\mathrm{FP}(\omega) = \sum_\mathbf{k}S_\mathbf{k}\delta(\omega-\omega_\mathbf{k})$, captures the phonons calculated from first principles, where $S_\mathbf{k}$ is the partial Huang-Rhys factor associated with a phonon with frequency $\omega_\mathbf{k}$~\cite{kamper2022signatures}.
To move to a continuum limit version, we account for the natural lifetime of phonons in a material by approximating the $\delta$-functions in the definition of $\mathcal{J}_\mathrm{Ph}(\nu)$ with a Gaussian function $\delta(\omega)\approx [\sigma\sqrt{2\pi}]^{-1}\exp[-(\omega/\sqrt{2}\sigma)^2]$, where $\sigma$ is the phonon broadening parameter.
We choose the broadening parameter as $\sigma=5$~meV, such that it phenomenologically reproduces features in the emission spectrum at low temperature.\par
The second contribution to the phonon spectral density, $J_\mathrm{A}(\omega)$, is associated to acoustic phonons, and leads to  the inhomogeneous broadening around the ZPL, which is not captured by the first-principle calculations.
We assume that this spectral density takes the form ${J}_\mathrm{A}(\omega) = \alpha \omega_c^{-2}\omega e^{-\omega^2/\omega_c^2}$~\cite{Nazir_PolaronMethodReview2016}, where $\alpha$ and $\omega_c$ are left as fitting parameters. This form of spectral density is commonly used in the semiconductor quantum dot literature, where it describes bulk acoustic phonons.
Similarly to the ZPL fits we restrict ourselves to a frequency window of $-0.06~\mathrm{eV}< \Delta\omega<0.01~\mathrm{eV}$. 
In contrast to the ZPL fits, however, we use the full expression for the emission spectrum, including the full phonon sideband given in Section~\ref{sec:SI_corr}.
The value of the optimal phonon parameters extracted from these fits changes depending on the defect.
It was not possible to obtain a fit for the acoustic phonon sideband for V$_\mathrm{N}$C$_\mathrm{B}$ due to the discrepancy between the sideband produced by first principles calculations and the measured PL. For illustrative reasons, we use the acoustic phonon parameters associated to C$_{2}$C$_\mathrm{N}$ when plotting the theoretical V$_\mathrm{N}$C$_\mathrm{B}$ PL.\par

\begin{figure}
    \centering
    \includegraphics[width=\columnwidth]{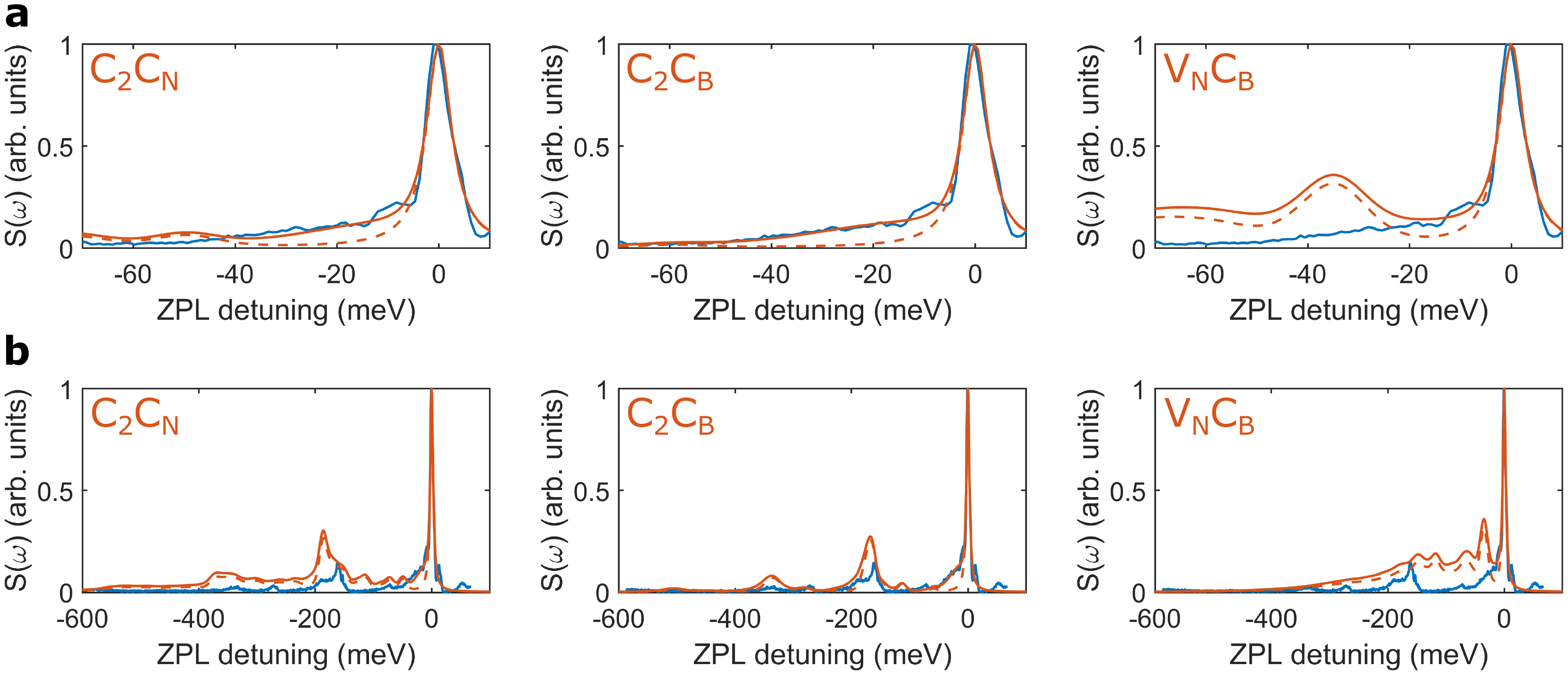}
    \caption{Comparison of the experiment (blue) and theory fits for three main defects considered in the main text. The dashed curves include only phonons calculated through \emph{ab initio} calculations, while the solid orange curves show the fits with acoustic phonons. The spectra plotted in (b,d,f) are the same as (a,c,e) respectively, plotted over a larger range. Temperature is set to $T=10$~K.}
    \label{fig:SI_fits}
\end{figure}

Fig.~\ref{fig:SI_fits} shows the fitted PL lineshapes for the three defects considered in the manuscript. We compare the cases where acoustic phonons are neglected (dashed) and included (solid). 
For the C$_{2}$C$_\mathrm{N}$ and C$_{2}$C$_\mathrm{B}$ defect complexes, we see improved agreement about the ZPL when acoustic phonons are included. V$_\mathrm{N}$C$_\mathrm{B}$ remains a are poor fit, with and without acoustic phonons.
Notably, when acoustic phonons are included, we observe additional acoustic sidebands on the phonon replicas found from DFT calculations.

\subsection{Theoretical photoluminescence emission spectroscopy}
The model presented above, and the resulting fits, allow us to calculate the PLE directly from the full phonon model.
We do this by varying the detuning of the continuous wave driving laser, $\Delta$, and calculating the emission spectrum for each detuning.
For the driven case, we consider the steady-state emission spectrum, defined as:
\begin{equation}
    S(\omega) = \mathrm{Re}\left[
    \lim\limits_{t\rightarrow\infty}\int\limits_0^\infty d\tau \mathcal{G}(\tau)g^{(1)}_\mathrm{opt}(t, \tau)
    \right],
\end{equation}
where the integral over $t$ has been reduced to the steady state limit.\par

\begin{figure}[h]
    \centering
    \includegraphics[width=\columnwidth]{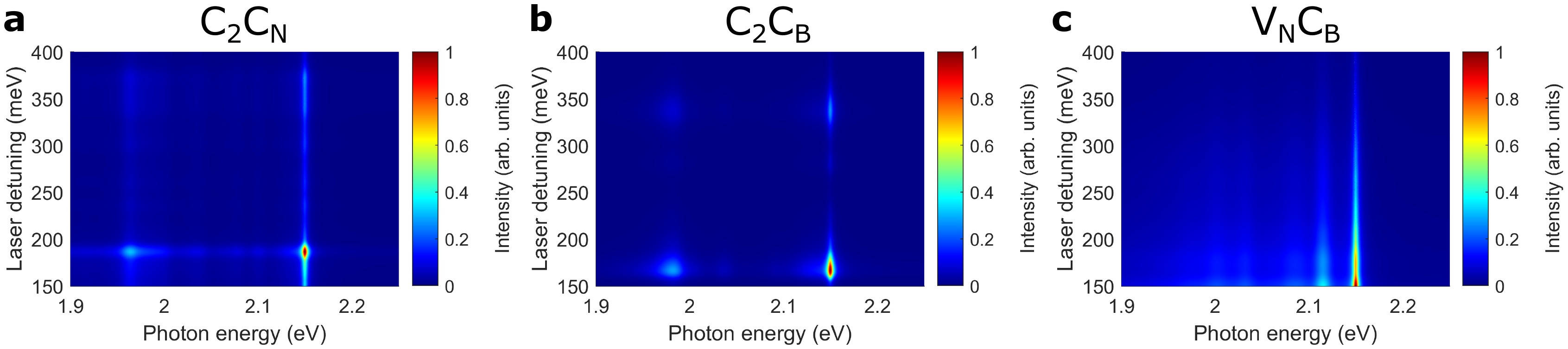}
    \caption{Theoretical PLE maps fitted to Emitter~A. The calculated PLE maps include electron-phonon interactions as described in the main text.}
    \label{fig:SI_PLE}
\end{figure}

Fig.~\ref{fig:SI_PLE} shows the PLE for the three defects considered in the main manuscript.

\begin{figure}[h]
    \centering
    \includegraphics[width=0.35\columnwidth]{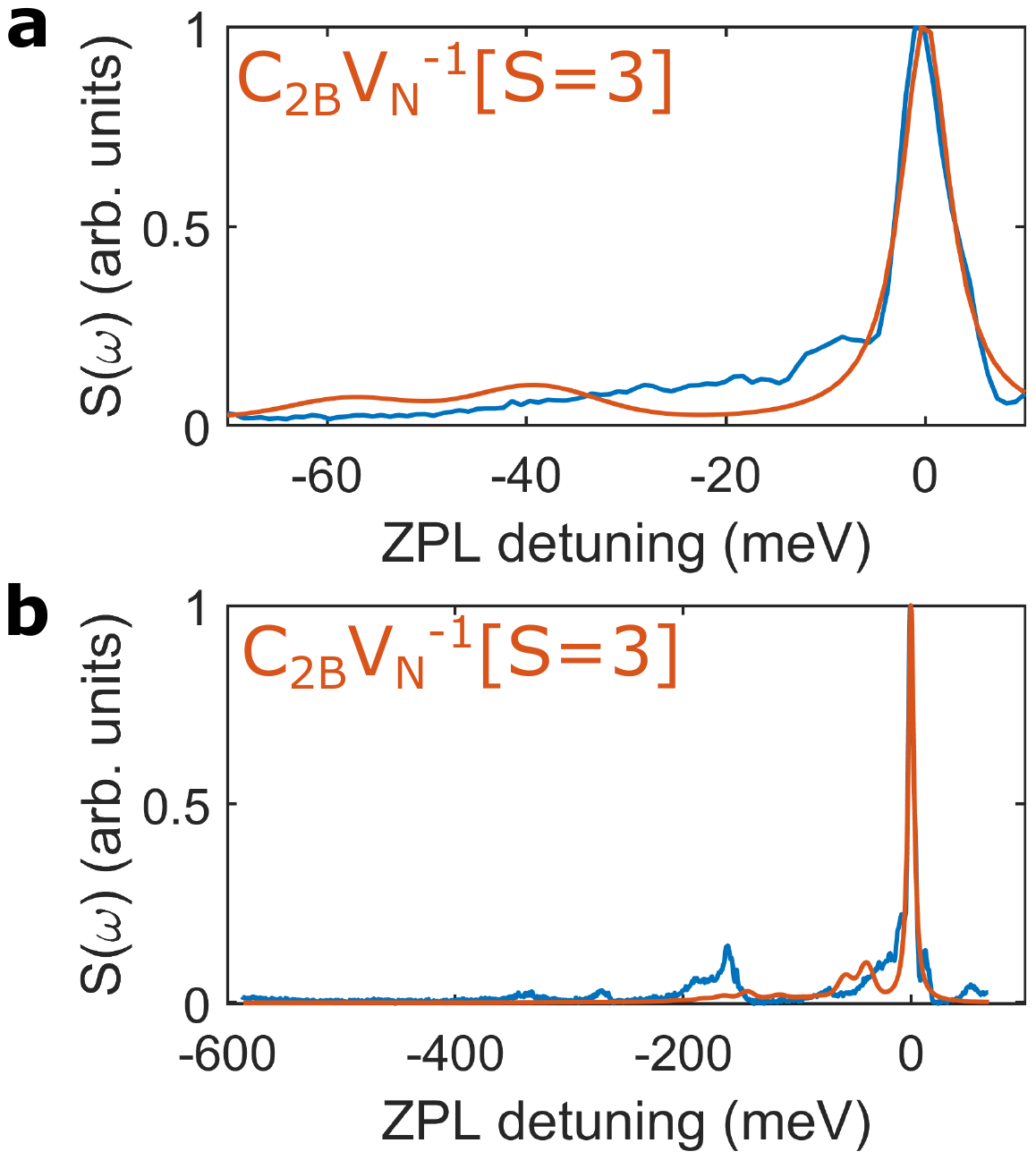}
    \caption{Comparison of the experimental photoluminescence (blue) with the theoretical spectrum of $\mathrm{C_{2B}V_N^{-1}[S=3]}$ (red). The theoretical spectrum is calculated without acoustic phonons. The spectra in~(a) and~(b) are identical but plotted over different ranges.}
    \label{fig:SI_c2bvnm1}
\end{figure}

\newpage
Fig.~\ref{fig:SI_c2bvnm1} shows the photoluminescence for $\mathrm{C_{2B}V_N^{-1}[S=3]}$ which does not conform nicely with the experiment.

\newpage
\section{Detailed scatter plots}\label{subsec:DetailedScatterPlots}

\begin{figure*}[h]
    \centering
    \includegraphics[width=\textwidth]{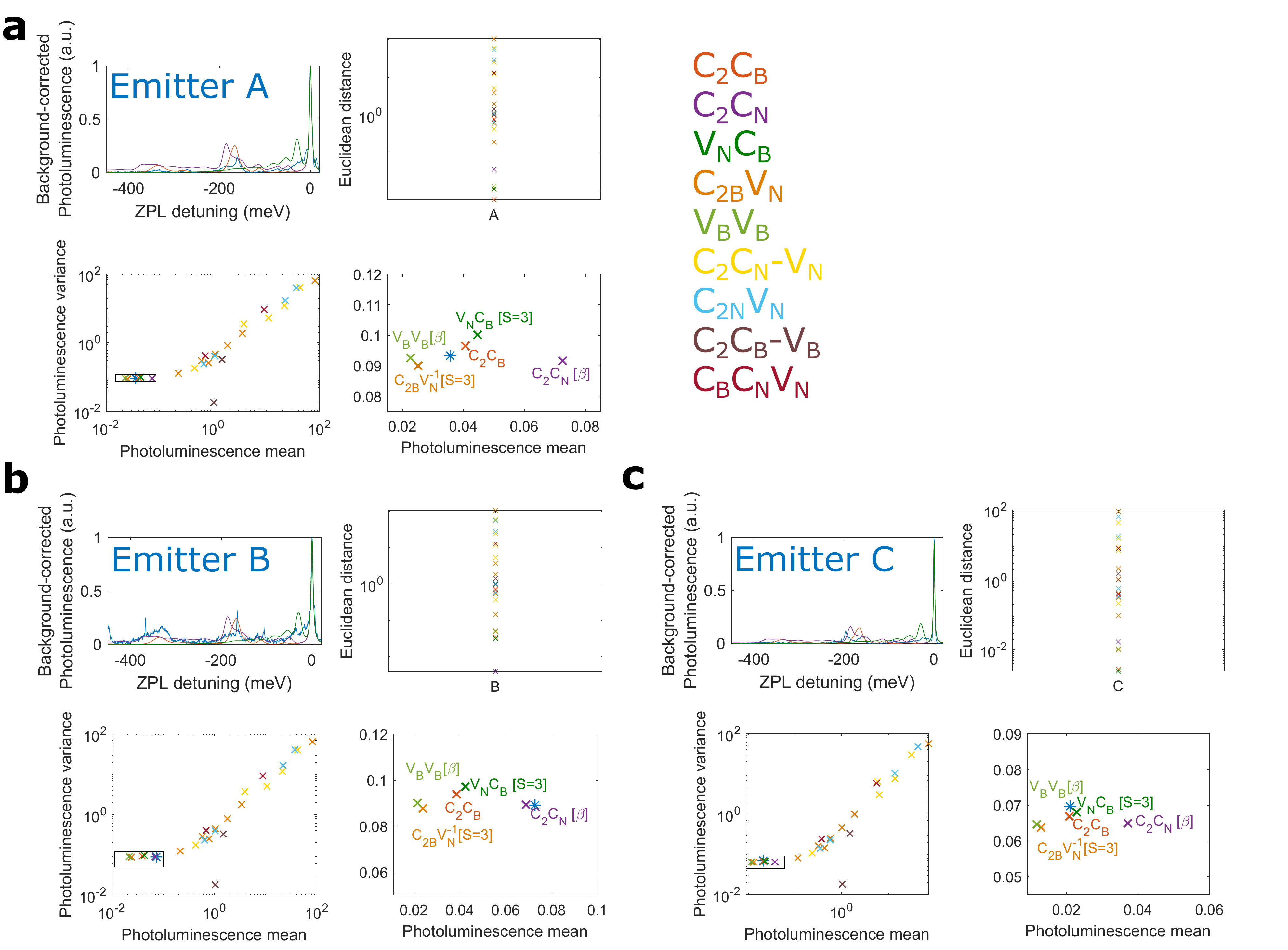}
    \caption{\textbf{Detailed scatter plot for all studied defects.} (a)~Emitter~A. The top left panel shows the experimental photoluminescence as well as the theoretical line shapes of $\mathrm{C_2C_B}[\beta]$ (red), $\mathrm{C_2C_N}$ (purple), and $\mathrm{V_NC_N[S=3]}$ (dark green). The bottom panels show the scatter plot with a zoom-in on the right side, indicated by a square in the left plot. The top right panel shows the euclidean distance. The colors of the defects are given on the right. (b)~Emitter~B, labelling as in~(a). (c)~Emitter~C, labelling as in~(a).}
\end{figure*}

\newpage
\section{Background correction and further low-temperature PSB}\label{subsec:SI_BackgroundAndPSB}
\begin{figure}[h]
  \includegraphics[width=0.5\columnwidth]{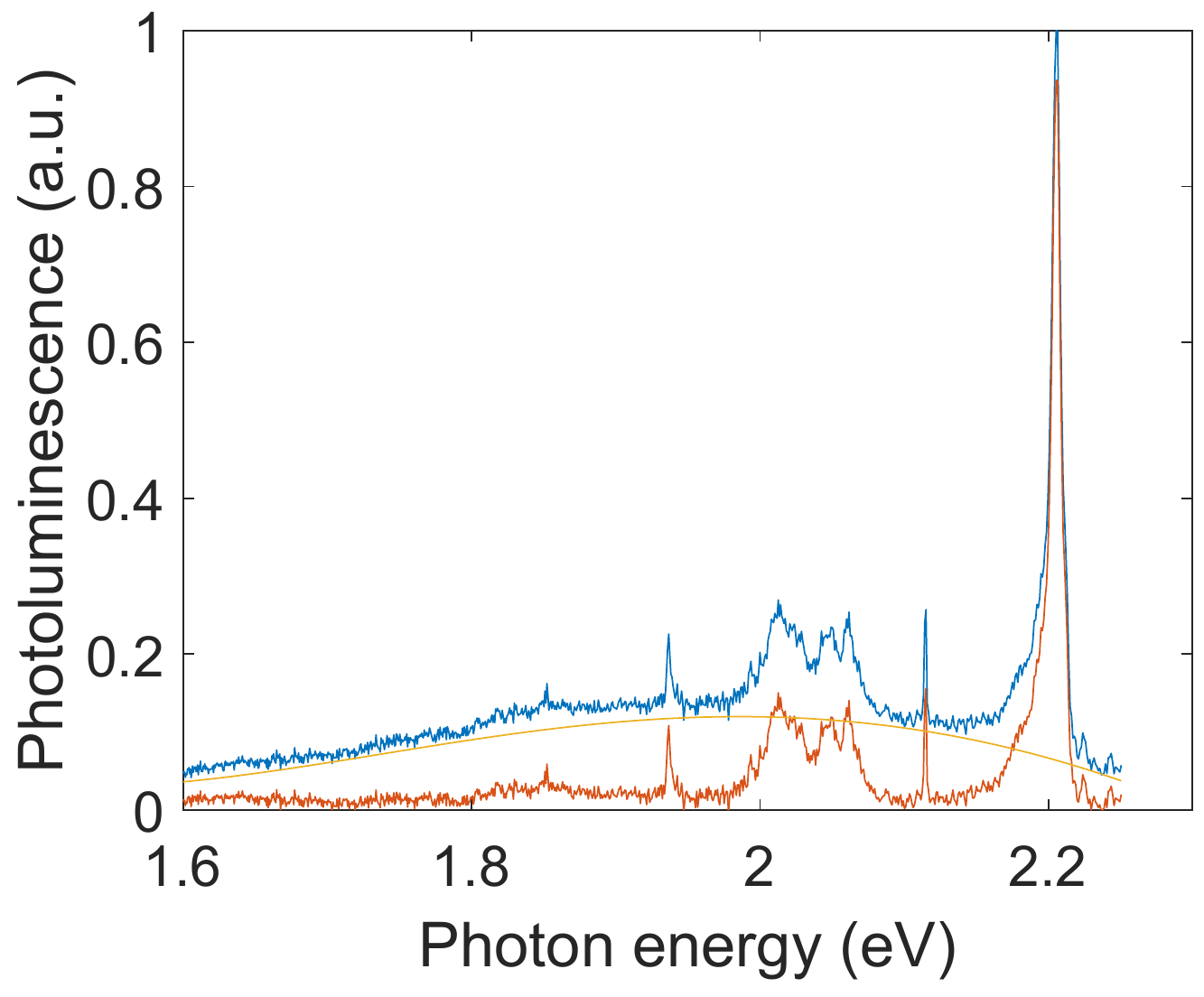}
  \caption{\label{figSI:BackgroundAndPSB}\textbf{Background correction of photoluminescence line shapes.}~Background correction of one photoluminescence spectrum. The raw spectrum is shown in blue, the background in yellow and the background-corrected spectrum in red.}
\end{figure}

For easy comparison of PSB, we carried out a background correction via a publicly available script~\cite{BackCorMatlab}. We use this script with the asymmetric Huber function at fourth order and $s=\SI{0}{}$. In Fig.~\ref{figSI:BackgroundAndPSB}, we show the background correction for one photoluminescence spectrum.

\newpage
\section{Extended level mechanism of $\mathbf{V_NC_B}$}\label{subsec:SI_Fig3_VNCBLevelScheme}

\begin{figure*}[h]
    \centering
    \includegraphics[width=0.4\textwidth]{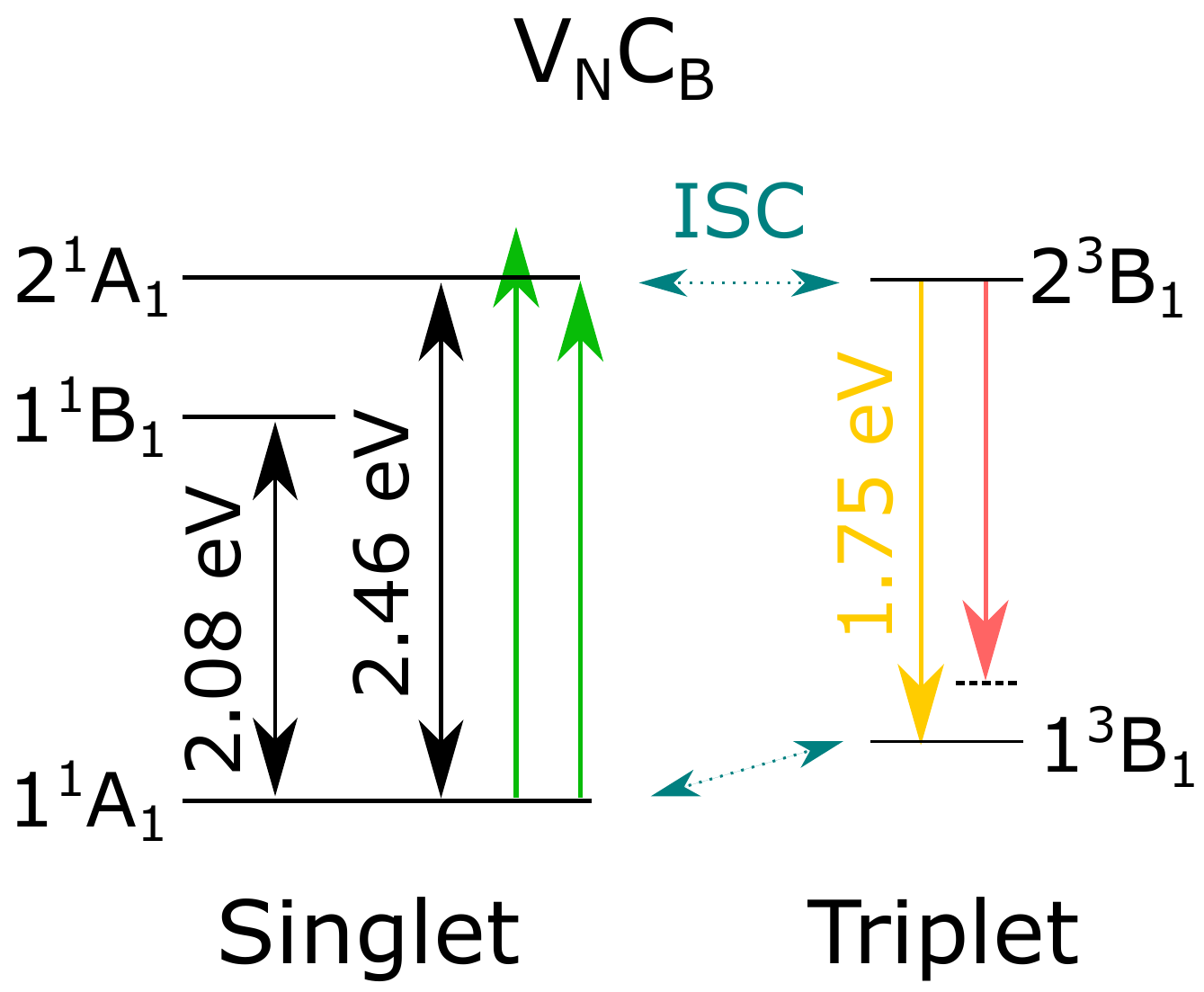}
    \caption{\textbf{Extended level mechanism of $\mathbf{V_NC_B}$.} The singlet system is shown on the left and the triplet system on the right. Emission from the triplet state is not possible by direct excitation because the ground state of the triplet state is not populated in equilibrium. However, optical excitation of the singlet system followed by an intersystem crossing (ISC) is possible. This requires laser detunings that are higher than $\SI{2.46}{} - \SI{1.75}{eV} = \SI{0.71}{eV}$. This is significantly larger than our experimental laser detunings from $\sim \SI{150}{}$ to $\SI{400}{meV}$.}
\end{figure*}

\newpage
\section{Further experimental PLE data}\label{subsec:SI_PLEDetails}

\begin{figure*}[h]
  \includegraphics[width=0.95\textwidth]{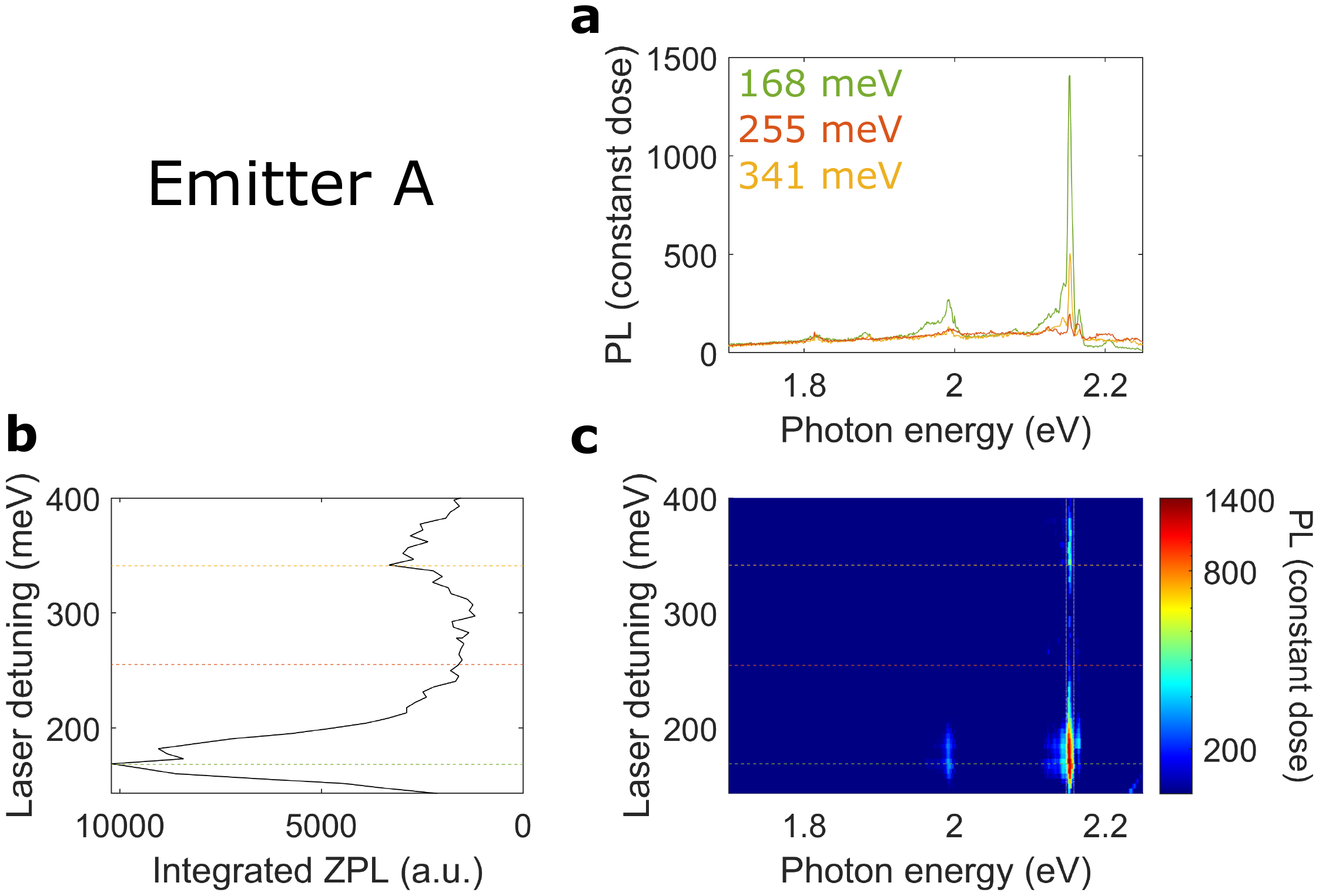}
  \caption{\label{figSI:PLEDetails}\textbf{PLE of Emitter~A at 10~K.}
  (a)~Photoluminescence spectra of the luminescent centre at several laser detunings, given by the legend. (b)~ZPL intensity as a function of laser detuning. The spectral range for the ZPL intensity is shown by the vertical lines in (c). (c)~PLE map of a luminescent centre in hBN. The horizontal lines correspond to the spectra shown in (b) and the dim upward line in the bottom right corner corresponds to the silicon Raman $\sim \SI{520}{cm^{-1}}$.}
\end{figure*}

\begin{figure*}[h]
    \includegraphics[width=0.7\textwidth]{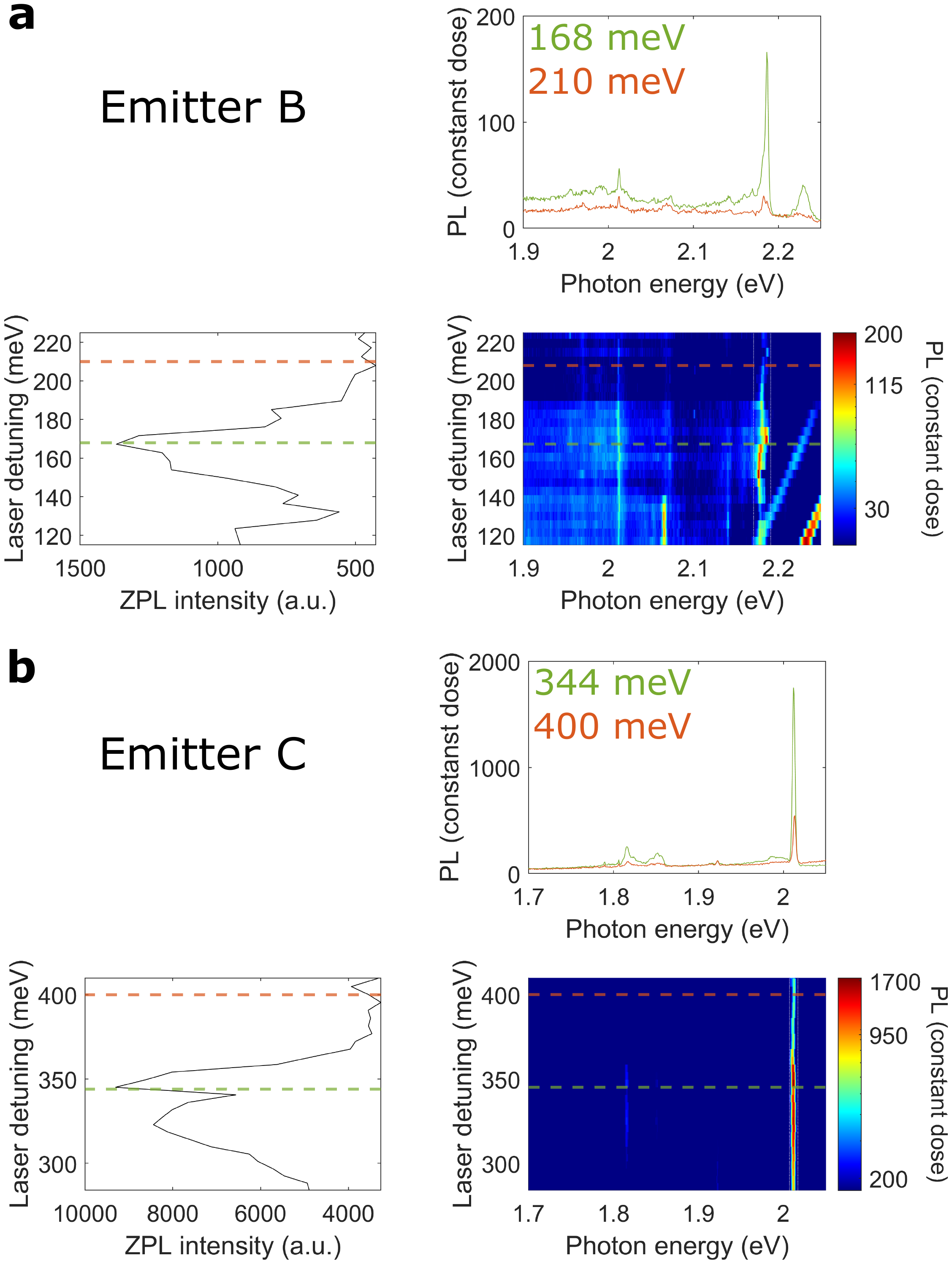}
    \caption{\label{figSI:PLE_EmitterBC}\textbf{PLE of further luminescent centres at $\SI{10}{K}$.} (a)~PLE measurement of Emitter~B, arranged and labelled as in Fig.~\ref{figSI:PLEDetails}. (b)~PLE measurement of Emitter~C, arranged and labelled as in Fig.~\ref{figSI:PLEDetails}.}
\end{figure*}

Fig.~\ref{figSI:PLEDetails} and~\ref{figSI:PLE_EmitterBC} show the PLE data for Emitter~A,~B and~C without any background correction, i.e. both photoluminescence and PLE intensities are shown without background correction.\par
Fig.~\ref{figSI:PLEDetails}a shows the photoluminescence spectra for several laser detunings. We obtain strong variations with laser detuning that are investigated in detail under continuous laser detuning, as shown in Fig.~\ref{figSI:PLEDetails}c. To investigate the absorption characteristic, we study the ZPL intensity which is defined as the area under the ZPL with a bandwidth of $\SI{10}{meV}$~(Fig.~\ref{figSI:PLEDetails}b). We observe a strong ZPL intensity at a detuning of $\SI{168}{meV}$ while an intermediate ZPL intensity is obtained at a detuning of $\SI{341}{meV}$. These two important detunings are highlighted by horizontal lines in Fig.~\ref{figSI:PLEDetails}c and the corresponding photoluminescence spectra are shown in Fig.~\ref{figSI:PLEDetails}a. Another important detuning is exactly between the two aforementioned, i.e. at $\SI{255}{meV}$, which yields a small ZPL intensity because it lays between the two states $\ket{e,1}$ and $\ket{e,2}$. 

\newpage

\newpage
\section{Rescaling of PLE spectra for Emitter A}\label{subsec:SI_RescalingPLE}
\begin{figure*}[h]
  \includegraphics[width=0.95\textwidth]{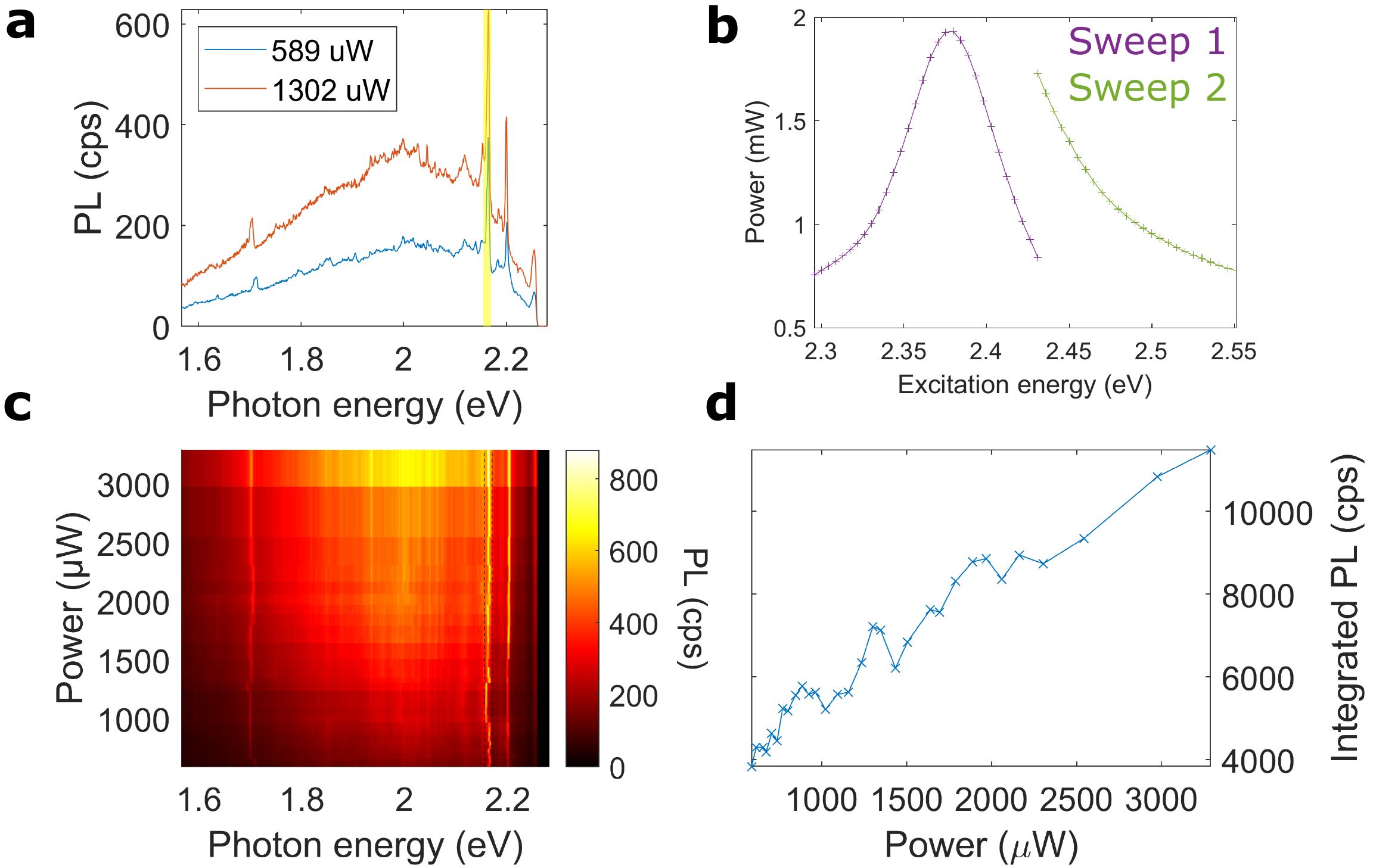}
  \caption{\label{figSI:Power}\textbf{Power dependence of excitation source and photoluminescence.} (a)~Photoluminesence spectra at several excitation powers, given by the legend. (b)~The excitation power as a function of excitation energy. We performed two laser sweeps with different laser settings to obtain a similar power for all laser detunings. (c)~Power-dependent photoluminescence. cps, counts per second. (d)~Integrated zero-phonon line (ZPL) as a function of excitation power (not background corrected). The integrated spectral range ($\SI{2.155}{}$ to $\SI{2.170}{eV}$) is shown by vertical dashed lines in~(c). We observe a linear dependence of the ZPL intensity as a function of excitation power.}
\end{figure*}

The PLE measurements shown in Fig.~2a 
are done by two laser sweeps: Sweep~1 and Sweep~2. We use two different laser settings to keep the power between $\SI{0.75}{}$ and $\SI{1.93}{mW}$ (see Fig.~\ref{figSI:Power}a). We first describe the rescaling within one sweep, followed by the rescaling between two sweeps. We finally describe the rescaling to one excitation power.\par
We measure the excitation power as a function of excitation energy prior to the PLE measurements (Fig.~\ref{figSI:Power}a). Within one sweep, the exposure time at power $P_i$ is set to
\begin{align}
    t_1\cdot\frac{P_1}{P_i}~,
\label{SIeq:rescale1}
\end{align}
where $t_1$ and $P_1$ are the exposure time and power at the beginning of the sweep while $P_i$ is the current power (at a specific laser energy or laser detuning). This adjustment in exposure time is meaningful, since the luminescent centres studied show a linear dependence of the intensity with excitation power in the studied power range, as shown in Fig.~\ref{figSI:Power}d. Furthermore, the detector operates in a constant-count regime which avoids saturation.\par
At each laser detuning in Sweep~2, the exposure time is adjusted as described by equation (\ref{SIeq:rescale1}). To rescale the spectra of Sweep~2 to Sweep~1, we multiply each spectrum of Sweep~2 (after adjusting the exposure time) by
\begin{align}
    \frac{t_1}{t_1^*}\cdot \frac{P_1}{P_1^*}~,
\label{SIeq:rescale2}
\end{align}
where $t_1^*$ and $P_1^*$ are the exposure time and the excitation power at the beginning of Sweep~2.\par
We will give an example why this multiplication is meaningful. At the end of Sweep~2, the excitation energy is identical to the beginning of Sweep~1. However, the excitation power at the end of Sweep~2 is $P_\text{end}^*$. Now we calculate the exposure time at the end of Sweep~2. Calculating the exposure time of Sweep~2, we multiply equation (\ref{SIeq:rescale1}) with equation (\ref{SIeq:rescale2}) and get
\begin{align}
    t_1^*\cdot\frac{P_1^*}{P_\text{end}^*}~\cdot~\frac{t_1}{t_1^*}\cdot \frac{P_1}{P_1^*} = t_1 \cdot \frac{P_1}{P_\text{end}^*}~
\end{align}
which is identical to the exposure time given by equation (\ref{SIeq:rescale1}) where $P_i$ is replaced by $P_\text{end}^*$.\par
After carrying out all the rescaling mentioned above, we rescale the data to $\SI{1}{mW}$ at an excitation energy of $\SI{2.431}{eV}$. This is the reason why we have shown absolute values in the photoluminescence (i.e. constant dose) in Fig.~2.
\newpage

\newpage
\section{PLE of optical and acoustic PSB of Emitter A}\label{subsec:SI_Fig4ExtendedDetunings}

\begin{figure*}[h]
  \includegraphics[width=0.8\textwidth]{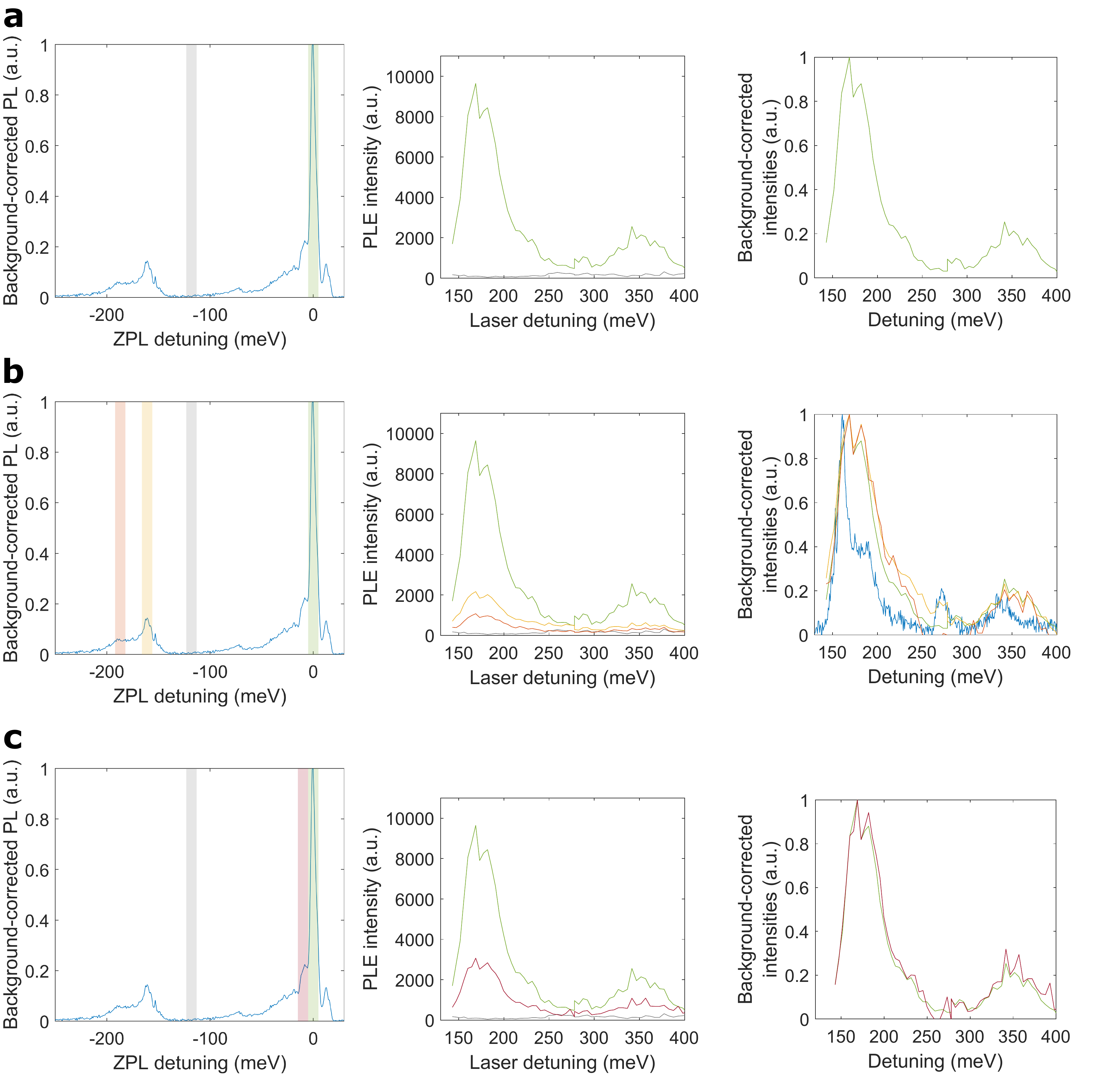}
  \caption{\label{figSI:Fig4ExtendedDetunings}\textbf{Background-correction of PLE intensities and acoustic PSB.} (a)~Background-correction of the ZPL intensity. The spectral ranges of ZPL and background are shown in the left panel, on top of the photoluminescence spectrum. The resulting PLE intensities are shown in the central panel and the background-corrected ZPL intensity in the right panel. The latter is defined as the green ZPL intensity minus the grey background intensity. The photoluminescence spectrum in~(a) is taken at a detuning of $\SI{168}{meV}$. (b)~Optical PSB intensities, labelled as in~(a). In the right panel, the flipped photoluminescence spectrum is shown in blue. (c)~Low-energy PSB intensity, labelling as in~(a).  }
\end{figure*}

Fig.~\ref{figSI:Fig4ExtendedDetunings}a shows how background-corrected PLE intensities are obtained, by using the ZPL intensity as an example. We use an unrelated photoluminescence range with the same width as the ZPL to define a background intensity (grey curve in the central panel). To obtain the background-corrected ZPL intensity, we subtract this background intensity. The resulting background-corrected ZPL intensity is shown in the right panel of Fig.~\ref{figSI:Fig4ExtendedDetunings}a. Similarly, we obtain the background-corrected intensities for several PSB, as shown in Fig.~\ref{figSI:Fig4ExtendedDetunings}b and in the main text.\par
Fig.~\ref{figSI:Fig4ExtendedDetunings}b shows the optical PSB intensities. The right panel shows that the optical PSB intensities at one-phonon detuning ($\SI{170}{meV}$) and the ZPL intensity at two-phonon detuning ($\SI{340}{meV}$) show comparable strength. All these processes are two-phonon processes (see Methods of the main text).\par
Fig.~\ref{figSI:Fig4ExtendedDetunings}c shows the low-energy PSB at detunings around $\SI{10}{meV}$. The low-energy PSB intensity is lower than for the ZPL, since it is a two-phonon process (see Methods and discussion in the main text). For the background-corrected intensities, we observe an almost identical qualitative behaviour of the low-energy PSB and ZPL intensity.
This supports our interpretation that the low-energy PSB and the ZPL have the same excitation mechanism.\par

\newpage
\section{Experimental line shapes different to group~I}\label{subsec:NoGroupISpectra}

\begin{figure}[h]
  \includegraphics[width=0.8\columnwidth]{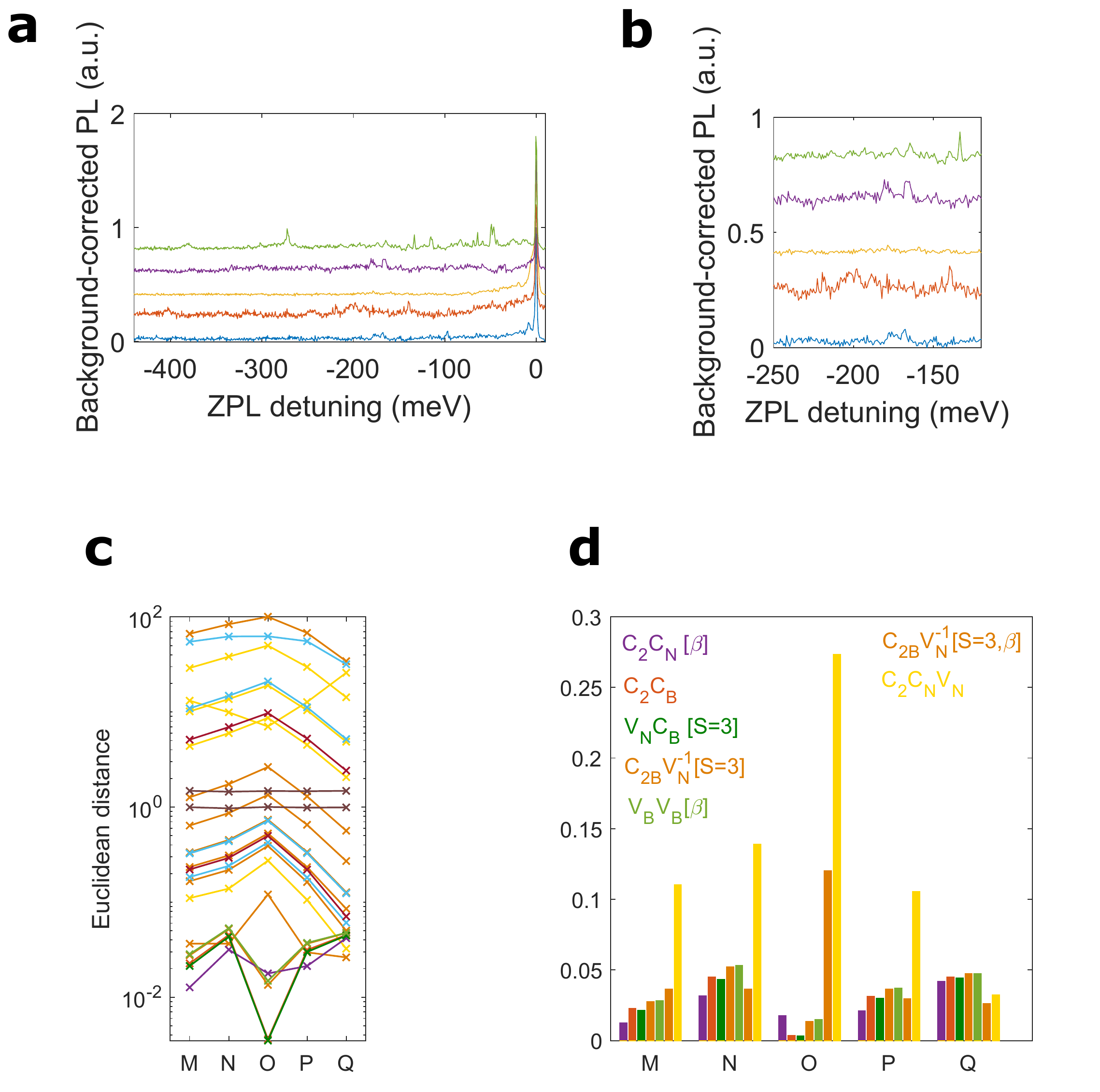}
  \caption{\textbf{Low-temperature photoluminescence of luminescent centres not belonging to group~I.} (a)~Photoluminescence line shapes at $T = \SI{10}{K}$ under $\SI{2.37}{eV}$ excitation, shifted vertically for clarity. The spectra from bottom to top correspond to Emitter~M to Emitter~Q, given by the x axis in~(c). (b)~Optical PSB of the luminescent centres shown in (a) with corresponding colors, shifted vertically for clarity.
  (c)~Euclidean distance for Emitter~M up to Emitter~Q, with corresponding spectra shown in~(a) from bottom to top. (d)~Histograms of Emitter~M to Emitter~Q. The colors correspond to the defect transitions given by the legend.  
  All line shapes are background corrected as outlined in Supplementary Information~\ref{subsec:SI_BackgroundAndPSB} and more details are given in the main text.}
\end{figure}
\begin{table}[h]
    \centering
    \begin{tabular}{c|c|c|c|c}
         M  &   N   &   O   &   P   &   Q   \\\hline
    2.0696  &2.0139 &2.1883 &2.1210 &2.1199
    \end{tabular}
    \caption{Zero-phonon line energies (in electron-volt) of Emitter~M to~Q.}
\end{table}

\newpage
\section{Optical characterisation}\label{subsec:SI_Optical characterisation}
For optical characterisation, we use an objective (details below) to focus a laser onto the sample. For photoluminescence (PL), we use a continuous-wave (CW) laser with a photon energy of $\SI{2.37}{eV}$. 
A helium flow cryostat is used for the low temperature measurements.\par
For PLE measurements at $\SI{10}{K}$, we use a supercontinuum white light laser ($\SI{78}{MHz}$ repetition rate) 
which is filtered down to a bandwidth of $\sim \SI{1}{nm}$ with a tunable laser line filter. 
In order to compensate for intensity changes at different excitation wavelengths, we adjust the integration time of the spectrometer (see Supplementary Information~\ref{subsec:SI_RescalingPLE}). For both photoluminescence (PL) and PLE measurements, the collected light is filtered by a longpass dichroic mirror ($\SI{550}{nm}$) 
and a $\SI{550}{nm}$ longpass filter. 
The filtered light is focused on the input slit of a spectrometer where the spectrum is obtained. The grating and detector information are given below.\par
For room temperature PL measurements, a 50X objective ($\text{NA} = \SI{0.6}{}$), a $\SI{150}{\text{lines/mm}}$ grating, and an EMCCD 
are used. The low-temperature PL and PLE experiments are performed using a home-made confocal microscope with a 50X objective ($\text{NA} = \SI{0.42}{}$), a $\SI{500}{\text{lines/mm}}$ grating, and a CCD detector. 
Furthermore, a spatial filter is used in front of the spectrometer for low-temperature measurements.

\newpage
\section{Modelling phonon effects in photoluminescence emission from luminescence centers}\label{subsec:JakeModel}

\subsection{The model Hamiltonian}
We model the studied defect complex as a two level system (TLS) with ground state $\ket{g}$ and excited state $\ket{e}$, with splitting $\omega_e$ ($\hbar=1$). 
The TLS is driven by a continuous wave laser with a frequency $\omega_L$ and Rabi coupling $\Omega$. 
The emitter couples to both a vibrational and optical environment, characterised by the Hamiltonian~\cite{Nazir_PolaronMethodReview2016}:
\begin{equation}
H(t) = \omega_e\ket{e}\!\bra{e} + \Omega\cos(\omega_L t)\sigma_x + \ket{e}\!\bra{e}\sum_\mathbf{k} g_k\left(b_\mathbf{k}^\dagger + b_{-\mathbf{k}}\right) + \sigma_x\sum_l \left( h^\ast_la_l^\dagger +h_l  a_l\right) + \sum_l \omega_l a_l^\dagger a_l  + \sum_\mathbf{k}\nu_k b^\dagger_\mathbf{k} b_\mathbf{k},
\end{equation} 
where we have defined $a_l^\dagger$ as the creation operator for a photon with energy $\omega_l$, and $b_\mathbf{k}^\dagger$ as the creation operator for a phonon with energy $\nu_\mathbf{k}$ and wavevector $\mathbf{k}$.
We have also introduced the system operators $\sigma_x = \sigma^\dagger + \sigma$ and $\sigma = \ket{g}\!\bra{e}$.
The coupling to the vibrational and electromagnetic environments are characterised by their respective spectral densities: for the optical environment we make the standard quantum optics approximation that the coupling constants $h_l$ are frequency independent, such that $J_\mathrm{EM}(\omega)\approx\frac{\Gamma}{2\pi}$, where $\Gamma$ is the emission rate of the optical transition; 
the phonon spectral density takes the form $J_\mathrm{Ph}(\nu) = \sum_\mathbf{k} S_\mathbf{k}\delta(\nu- \nu_\mathbf{k})$ where $S_\mathbf{k} = \omega_\mathbf{k}^2\vert g_\mathbf{k}\vert^2$ are the partial Huang-Rhys parameters, and contain contributions from phonons calculated through \emph{ab initio} methods and those fitted to the emission as detailed below.\par

We can simplify the above equation by making the rotating-wave approximation and moving to a frame rotating with respect to the laser frequency $\omega_L$, yielding the Hamiltonian:
\begin{equation}
H = \Delta\ket{e}\!\bra{e} +\frac{\Omega}{2}\sigma_x + \ket{e}\!\bra{e}\sum_k g_k (b_k^\dagger + b_k) + \sum_l \left(h^\ast_l\sigma a_l^\dagger e^{i\omega_L t} +h_l  \sigma^\dagger a_l e^{-i\omega_L t}\right)
+ \sum_l \omega_l a_l^\dagger a_l  + \sum_k\nu_k b^\dagger_k b_k,
  \end{equation}
where $\Delta = \omega_e- \omega_L$ is the detuning between the driving field and the exciton transition. 
This Hamiltonian forms the starting point of our analysis of the dynamical and optical properties of the defect.

\subsection{Polaron theory for a driven emitter }
In order to account for the strong coupling to the vibrational environment, we apply a polaron transformation to the global Hamiltonian, i.e. the unitary transformation $\mathcal{U}_P = \ket{e}\!\bra{e}\otimes B_+  + \ket{g}\!\bra{g}$, where $B_\pm = \exp\left(\pm\sum_\mathbf{k} \nu_\mathbf{k}^{-1} g_k(b_\mathbf{k}^\dagger - b_{-\mathbf{k}})\right)$ are displacement operators of the phonon environment~\cite{mccutcheon2010quantum,Nazir_PolaronMethodReview2016,iles2017limits}. This transformation dresses the excitonic states with vibrational modes of the phonon environment.
In the polaron frame, the Hamiltonian may be written as $H_P= H_0  + H^\mathrm{PH}_\mathrm{I}+ H^\mathrm{EM}_\mathrm{I}$, where  
\begin{equation}
\begin{split}
H_0& = \tilde\Delta\ket{e}\!\bra{e} + \frac{\Omega_R}{2}\sigma_x + \sum_l \omega_l a_l^\dagger a_l  + \sum_k\nu_k b^\dagger_k b_k,\\
H^\mathrm{PH}_\mathrm{I}& = \frac{\Omega}{2}\left(\sigma_x B_x + \sigma_y B_y\right),\hspace{0.5cm}\mathrm{and}\hspace{0.5cm} H^\mathrm{EM}_\mathrm{I}  = \sum_l h_l ^\ast \sigma B_+ a_l^\dagger e^{i\omega_L t}+ \mathrm{h.c.},
\end{split}
\end{equation}
where we have introduced the phonon operators $B_x = (B_+ + B_- - 2B)/2$, $B_y = i(B_+ - B_-)/2$, and the Frank-Condon factor of the phonon environment $B = \text{tr}_B(B_\pm\rho_B) = \exp(-(1/2)\sum_\mathbf{k} \nu_\mathbf{k}^{-2}\vert g_\mathbf{k}\vert^2\coth(\nu_\mathbf{k}/k_BT))$, with the Gibbs state of the phonon environment in the polaron frame given by $\rho_B = \exp(-\sum_\mathbf{k}\nu_\mathbf{k} b_\mathbf{k}^\dagger b_\mathbf{k}/k_BT)/ \text{tr}  \left(\exp(-\sum_\mathbf{k}\nu_\mathbf{k} b_\mathbf{k}^\dagger b_\mathbf{k}/k_BT)\right)$. 
Notice that the polaron transformation has dressed the operators in the light-matter coupling Hamiltonian, and renormalised the system parameters, such that $\Omega_R= \Omega B$ and $\tilde{\Delta} = \Delta - \sum_\mathbf{k} \nu_\mathbf{k}^{-1} g^2_\mathbf{k}$.\par

The dynamics of the electronic states of the emitter are described by a 2$^\textrm{nd}$-order Born-Markov master equation~\cite{breuer2002theory},  where both the electromagnetic field and the transformed vibrational environment are perturbatively eliminated. The polaron transformation means that the vibrational interaction is included non-perturbatively in the interaction strength~\cite{Nazir_PolaronMethodReview2016}. 
We can use the fact that only terms quadratic in field operators are non-zero when traced with a Gibbs state, so that there will be no cross-terms between the vibrational and electromagnetic dissipators, such that the master equation can be written in two parts:
\begin{equation}
\frac{\partial\rho(t)}{\partial t} = -i\left[\tilde\Delta\ket{e}\!\bra{e} + \frac{\Omega}{2}\sigma_x,\rho(t)\right]+ \mathcal{K}_\mathrm{PH}[\rho(t)] + \frac{\Gamma}{2}L_\sigma[\rho(t)] + \frac{\gamma}{2}L_{\sigma^\dagger\sigma},
\label{eq:meq}
\end{equation}
where $\rho(t)$ is the reduced state of the emitter, $L_\sigma[\rho] = 2\sigma\rho\sigma^\dagger - \{\sigma^\dagger\sigma,\rho\}$ is the Lindblad form dissipator for the electromagnetic environment, and similarly $\mathcal{K}_\mathrm{PH}$ describes the dissipation due to the vibrational environment.
Note that we have also included a source of homogeneous broadening given by rate $\gamma$, to account for experimental imperfections and other dephasing mechanisms such as charge noise.
For the optical dissipator, we refer the reader to standard texts in quantum optics for the derivation of the optical component~\cite{breuer2002theory}, in the following section we outline the derivation of the phonon dissipator.\par

\subsubsection{Deriving the phonon dissipator}
Considering only the coupling to the vibrational modes, we 
move into the interaction picture with respect to the Hamiltonian $H_0$, the interaction term becomes:
\begin{equation}
H_\mathrm{I}^\mathrm{PH}(t) = \frac{\Omega}{2} [\sigma_x(t) B_x(t) + \sigma_y(t) B_y(t)],
\end{equation}
where the interaction picture system operators are given by:
\begin{equation}\begin{split}
\sigma_x(t)& = \eta^{-2}\left[\tilde\Delta\Omega_R(1-\cos(\eta t))\sigma_z + (\Omega_R^2 + \tilde\Delta^2\cos(\eta t))\sigma_x + \tilde\Delta \eta\sin(\eta t)\sigma_y
\right],\\
\sigma_y(t)&= \eta^{-1}\left(\Omega_R\sin(\eta t)\sigma_z + \eta\cos(\eta t)\sigma_y - \tilde\Delta\sin(\eta t)\sigma_x\right),
\end{split}\end{equation}
and we have defined the generalised Rabi frequency $\eta = \sqrt{\tilde\Delta^2 + \Omega_R^2}$.
The interaction picture bath operators are found as $B_{x/y}(t)=e^{i H_0 t}B_{x/y}e^{-i H_0 t}$, with $B_{x/y}$ defined above.
In the Schr\"odinger picture and making the Born-Markov approximation in the polaron frame~\cite{breuer2002theory}, the dissipator describing the electron-phonon interaction is given by: 
\begin{equation}
\mathcal{K}_\mathrm{PH} [\rho(t)]  = -\frac{\Omega^2}{4}\int\limits_0^\infty \left([\sigma_x,\sigma_x(-\tau)\rho(t)]\Lambda_{xx}(\tau) + [\sigma_y,\sigma_y(-\tau)\rho(t)]\Lambda_{yy}(\tau) + \operatorname{h.c.} \right)d\tau,
\end{equation}
where we have and introduced the phonon bath correlation functions as 
$
\Lambda_{xx}(\tau) = \langle B_x(\tau)B_x\rangle = B^2(e^{\varphi(\tau)} + e^{-\varphi(\tau)} - 2)
$, 
$\Lambda_{yy}(\tau) =  \langle B_y(\tau)B_y\rangle  = B^2(e^{\varphi(\tau)}-e^{-\varphi(\tau)})$ {which quantify the response of the phonon environment to the exciton dynamics. Taking the continuum limit over the phonon modes, these correlation functions can be described in terms of the polaron frame propagator $\varphi(\tau) = \int_0^\infty d\nu \nu^{-2}J(\nu)[\coth(\nu/2k_\mathrm{B}T)\cos(\nu\tau)- i\sin(\nu\tau)]$.}
We can simplify the form of the master equation by evaluating the integrals over $\tau$, leading to:
\begin{equation}
\mathcal{K}_\mathrm{PH}[\rho_s(t)] = -\frac{\Omega^2}{4} \left([\sigma_x, \chi_x\rho_s(t)] + [\sigma_y, \chi_y\rho_s(t)] + \operatorname{h.c.}\right),
\end{equation}
where we have introduced the rate operators:
\begin{align}
\chi_x &= \int\limits_0^\infty \sigma_x(-\tau)\Lambda_{xx} (\tau) d\tau = \frac{1}{\eta^2}\left[\Delta\Omega_r(\Gamma^x_0 - \Gamma^x_c) \sigma_z + (\Omega_r^2 \Gamma^x_0+ \Delta^2\Gamma^x_c)\sigma_x + \Delta\eta \Gamma^x_s \sigma_y\right] ,\\
\chi_y & = \int\limits_0^\infty \sigma_y(-\tau)\Lambda_{yy} (\tau) d\tau = \frac{1}{\eta}\left[\Omega_r\Gamma^y_s \sigma_z - \Delta\Gamma^y_s\sigma_x + \eta \Gamma^y_c \sigma_y\right],
\end{align}
and defined the terms $\Gamma^a_0 = \int_0^\infty\Lambda_{aa}(\tau)d\tau$, $\Gamma^a_c = \int_0^\infty\Lambda_{aa}(\tau)\cos(\eta\tau)d\tau$, $\Gamma^a_s = \int_0^\infty \Lambda_{aa}(\tau)\sin(\eta\tau)d\tau$, with $a \in \{x,y\}$.
We may understand these terms as the rates at which transitions occur between the eigenstates of the system (i.e. the dressed states) induced by phonons. 
Qualitatively, $\Gamma_s^a$ and $\Gamma_c^a$ describe processes where a polaron is formed during excitation of the emitter, resulting in an exchange of energy with the environment. 
In contrast, $\Gamma_0^a$ leads to no net exchange of energy, inducing a pure dephasing type process.\par
Since we are operating in the polaron frame, the above master equation is non-perturbative in the electron-phonon coupling strength, capturing strong-coupling and non-Markovian influences in the lab frame, despite having made a Born Markov approximation. This provides us with a simple, intuitive, and computationally straight-forward method for describing exciton dynamics in a regime where a standard weak coupling master equation would break down~\cite{Nazir_PolaronMethodReview2016}.

\subsection{Correlation functions and spectra for a driven emitter}
\label{sec:SI_corr}
We are interested in understanding the impact that phonon coupling has on the PLE spectrum, which can be calculated from the  first-order correlation function $g^{(1)}(t,\tau)$.
As shown in Ref.~\cite{iles2017phonon}, the first-order correlation function in the polaron frame for the emitter is written in terms of the polaronic dipole operator $\underline{\bm\sigma} = \sigma B_+$, that is, the correlation function is $g^{(1)}_0(t,\tau) = \langle B_-(t+\tau)\sigma^\dagger(t+\tau) B_+(t)\sigma(t)\rangle$.
In general, $B_\pm$ are many-body displacement operators that do not commute with system operators. 
However, in the regime that the polaron master equation is valid at second-order, we can factor this correlation function in the Born approximation, such that $g^{(1)}_0(t,\tau) \approx \mathcal{G}(\tau)g^{(1)}_\mathrm{opt}(t,\tau)$.
There are two terms in this expression, the correlation function for the scattered field $g^{(1)}_\mathrm{opt}(t,\tau) = \langle\sigma^\dagger(t+\tau)\sigma(t)\rangle$ describing the purely electronic transitions, and the phonon correlation function $\mathcal{G}(\tau) = \langle B_-(\tau)B_+\rangle = B^2\exp(\varphi(\tau))$ which accounts for the relaxation of the phonon environment.
This factorisation allows us to split the emission spectrum into two components $S(\omega) =S_\mathrm{opt}(\omega) + S_\mathrm{SB}(\omega)$, where $S_\mathrm{opt}(\omega) =\mathrm{Re}[B^2\int_0^\infty d t\int_0^\infty g^{(1)}_\mathrm{opt}(t,\tau)e^{i\omega\tau}d\tau]$ corresponds to purely optical transitions, and $S_\mathrm{SB}(\omega) =\mathrm{Re}[\int_0^\infty d t\int_0^\infty (\mathcal{G}(\tau)-B^2)g^{(1)}_\mathrm{opt}(t,\tau)e^{i\omega\tau}d\tau]$ is attributed to non-Markovian phonon relaxation during the emission process, and leads to the emergence of a phonon sideband~\cite{brash2019light}.

\newpage
\bibliography{PLEBib.bib}

\end{document}